\documentclass[11pt,a4paper]{article}
\usepackage{jcappub}
\usepackage{rotating}
\usepackage{graphicx}
\usepackage{wrapfig}
\usepackage{epsfig}
\usepackage{amssymb}
\usepackage{textcomp}
\usepackage{wasysym}
\usepackage{lineno}
\usepackage[ansinew]{inputenc}
\usepackage{xspace}

\usepackage{siunitx}
\usepackage{verbatim} % for multiline comments, can be removed later

\newcommand{\be}{\begin{equation}}
\newcommand{\ee}{\end{equation}}
\newcommand{\bea}{\begin{equation}\begin{aligned}}
\newcommand{\eea}{\end{aligned}\end{equation}}

\newcommand{\TSysTotal}{\SI{8}{K}}
\newcommand{\TSysReceiver}{\SI{6}{K}\xspace}
\newcommand{\TSysBooster}{\SI{2}{K}}
\newcommand{\TActualBoosterTemp}{\SI{4}{K}}
\newcommand{\DetectionEfficiency}{80\%\xspace}
\newcommand{\EpsiolonLaAl}{\num{24}\xspace}
\newcommand{\EpsiolonSaphire}{\num{11.5}\xspace}
\newcommand{\TanDeltaLaAl}{\ensuremath{10^{-6}}} 
\newcommand{\TanDeltaSaphire}{\ensuremath{10^{-4}}} 
\newcommand{\DiskNumber}{80}

\newcommand{\MagneticField}{\SI{10}{T}\xspace} 
\newcommand{\MagnetBore}{\SI{1}{m^2}\xspace}
\newcommand{\MagnetLength}{\SI{200}{\centi\meter}\xspace}
\newcommand{\MagnetBsquareA}{\SI{100}{T^2.m^2}\xspace}

\newcommand{\BenchmarkPowerBoost}{$\sim \num{5e4}$\xspace}
\newcommand{\BenchmarkBoostWidth}{\SI{50}{\mega\hertz}}
\newcommand{\BenchmarkPower}{\SI{1e-22}{\watt}\xspace} %TOCHECK !

\newcommand{\MassRange}{\SIrange{40}{400}{\micro\electronvolt}\xspace}

\newcommand{\FrequencyRange}{\SIrange{10}{100}{\giga\hertz}\xspace}

\newcommand{\BroadbandMeasurmentTime}{$\sim 4$ days}

% ...

\begin{document}

%\begin{frontmatter}

\title{A new experimental approach to probe QCD 
axion dark matter in the mass range above 40\,$\boldsymbol{\mu}$eV} 
\author{\bf The MADMAX collaboration:}%\linebreak
\author[h]{\\ P.\,Brun,}
\author[a]{A. Caldwell}
\author[h]{L.\,Chevalier,}
\author[a]{G.\,Dvali}
\author[d]{P.\,Freire}
\author[c]{E.\,Garutti}
\author[d]{S.\,Heyminck}
\author[g]{J.\,Jochum}
\author[a]{S.\,Knirck}
\author[d]{M.\,Kramer}
\author[c]{C.\,Krieger}
\author[h]{T.\,Lasserre,}
\author[a]{C.\,Lee}
\author[a]{X.\,Li}
\author[b]{A.\,Lindner}
\author[a]{B.\,Majorovits}
\author[c]{S.\,Martens}
\author[c]{M.\,Matysek}
\author[a,*]{A.\,Millar}
\author[a]{G.\,Raffelt}
\author[e,a]{J.\,Redondo}
\author[a]{O.\,Reimann}
\author[b]{A.\,Ringwald}
\author[a]{K.\,Saikawa}
\author[b]{J.\,Schaffran}
\author[f]{A.\,Schmidt}
\author[c]{J.\,Sch\"utte-Engel}
\author[a]{F.\,Steffen}
\author[g]{C.\,Strandhagen}
\author[d]{G.\,Wieching}

\affiliation[a]{Max-Planck-Institut f\"ur Physik, Munich, Germany}
\affiliation[b]{Deutsches Elektronen Synchrotron, DESY, Hamburg, Germany}
\affiliation[c]{University of Hamburg, Hamburg, Germany}
\affiliation[d]{Max-Planck-Institut f\"ur Radioastronomie, Bonn, Germany}
\affiliation[e]{Departamento de F\'isica Te\'orica, University of Zaragoza, Spain}
\affiliation[f]{RWTH Aachen, Germany}
\affiliation[g]{Universit\"at T\"ubingen, Germany}
\affiliation[h]{CEA-IRFU, Saclay, France}
\affiliation[*]{now at Stockholm University, Sweden}

\abstract{
The axion emerges in extensions of the Standard Model that explain the absence of CP violation in the strong interactions. Simultaneously, it can provide naturally the cold dark matter in our universe. Several searches for axions and axion-like particles (ALPs) have constrained the corresponding parameter space over the last decades but no unambiguous hints of their existence have been found. The axion mass range below 1\,meV remains highly attractive and a well motivated region for dark matter axions. In this White Paper we present a description of a new experiment based on the concept of a dielectric haloscope for the direct search of dark matter axions in the mass range of \MassRange.  This MAgnetized Disk and Mirror Axion eXperiment (MADMAX) will consist of several parallel dielectric disks, which are placed in a strong magnetic field and with adjustable separations. This setting is expected to allow for an observable emission of axion induced electromagnetic waves at a frequency between \FrequencyRange corresponding to the axion mass.
}

\maketitle
%\begin{keyword}
%Axion, Peccei Quinn symmetry, Dark Matter
%%\PACS numbers ot be filled out23.40.-s \sep 24.10.Lx \sep 29.40.Gx \sep 29.40.Wk
%\end{keyword}
%
%\end{frontmatter}

%\setcounter{pageno}{0}

%\setcounter{page}{1}

\section{Introduction}

Axions are hypothetical low-mass bosons predicted by the Peccei--Quinn (PQ) mechanism, which explains the absence of CP-violating effects in quantum chromodynamics (QCD)~\cite{Peccei:1977hh,Peccei:1977ur,Weinberg:1977ma,Wilczek:1977pj}. Axions could also provide the cold dark matter (DM) of the universe~\cite{Preskill:1982cy,Abbott:1982af,Dine:1982ah} and as such are among the few particle candidates that simultaneously resolve two major problems in physics.

Assuming axions make up most of the DM in the universe, their mass $m_a$ is expected to be less than $\sim$\,meV (cf.~\cite{Sikivie:2006ni,Marsh:2015xka} and references therein). Mass values higher than $\sim 20$\,meV are excluded due to astrophysical constraints, see \cite{raffelt} for a review and 
\cite{Chang:2018rso,Giannotti:2017hny,RingwaldPDG} for updates and descriptions of recent anomalies. 
The existing experimental efforts for DM axion searches focus on an $m_a$ range below $\sim$\,40\,$\mu$eV. This is motivated by the realignment mechanism of the axion field providing the right amount of DM in scenarios in which the PQ symmetry is broken before inflation and never restored thereafter. 
Among these experimental efforts are microwave cavity searches~\cite{sikivie} such as ADMX~\cite{admx,Du:2018uak}, ORGAN~\cite{McAllister:2017lkb}, 
HAYSTAC~\cite{admx-hf,Kenany:2016tta,Zhong:2018rsr} or CULTASK~\cite{CULTASK}, which have begun to probe part of the axion parameter space.

In scenarios in which the PQ symmetry is broken after inflation, the realignment mechanism now along with decaying topological defects provides a cold DM axion density that matches the observed value if the axion mass $m_a$ is of the order of 100\,$\mu$eV \cite{1,2,3,4,nature,smash}.  
One recent attempt to improve the numerical simulations points to a more concrete mass value of $m_a\sim\,26\,\mu$eV~\cite{Klaer:2017ond} but still faces large theoretical uncertainties~\cite{Gorghetto:2018myk}.

We propose to search for QCD axion DM in the mass range around 100~$\mu$eV, using a dielectric haloscope~\cite{prl}. This concept makes use of the ``dish antenna'' idea~\cite{dish} and of additional signal enhancements possible by having multiple dielectric layers~\cite{dielectric}. The proposed MAgnetized Disk and Mirror Axion eXperiment (MADMAX) will consist of a mirror and about 100 dielectric disks each about 1~m$^2$ large with adjustable separations placed inside a homogeneous $10$~T strong magnetic dipole field.

This White Paper gives a summary of the principles upon which dielectric haloscopes are based, followed by a description of the first baseline design that could be used for the search of axions with mass in the range of \MassRange.  The results of measurements at a test setup are presented, which lead us to the conclusion that it should be realistic to build an experiment that can cover a large fraction of the parameter space including the unexplored one predicted for DM axions in the post inflationary PQ symmetry breaking scenario.

\section{Theoretical motivation}

\subsection{Strong CP problem}

In the standard model (SM) of particle physics, violation of CP in the strong interactions is controlled by just one parameter, the $\theta$ angle. This angle appears as the sum of two contributions with a-priori unrelated origins: the angle defining the vacuum of QCD, $\theta_{\rm QCD}$, and the common phase of the quark mass matrix, Arg\,Det\,$M_q$, related to the Yukawa couplings of the Higgs sector. Observable effects derive only from this combination. When we redefine quark fields to make their masses real, the phase appears as the coupling constant of the topological charge density operator of QCD, i.e., the SM Lagrangian contains a term
\be
\frac{\alpha_s}{8\pi} \theta\, \widetilde G^{\mu\nu}_{a} G_{\mu\nu a}   
\equiv \frac{\alpha_s}{2\pi} \theta\, {\bf E}_a\cdot {\bf B}_a 
\label{eq:ThetaTerm}
\ee
which violates parity, time-reversal and thus CP. In~(\ref{eq:ThetaTerm}) $G_{\mu\nu a}$ denotes the QCD field-strength tensor with $\widetilde G^{\mu\nu}_a \equiv \frac{1}{2} \epsilon^{\mu\nu\alpha\beta} G_{\alpha\beta a}$ being its dual, color index $a = 1,...,8$, and ${\bf E}_a$ and ${\bf B}_a$ the illustrative chromo electric and chromo magnetic fields, respectively.
Non-zero values of $\theta$ imply CP violating observables such as nuclear electric dipole moments but none of these effects have been observed to date. As a key example, the electric dipole of the neutron is predicted to be $d_n = 2.4 \times 10^{-16} \theta e$~cm~\cite{Pospelov:1999mv} but the most recent experiment \cite{Baker:2006ts} concluded that $|d_n|<2.9\times 10^{-26} e$~cm, so $\theta$ must be extremely tiny, $\theta< 1.3\times 10^{-10}$.  
This is an amazingly small upper limit, especially if we consider that the only other CP violating phase in the SM, the CKM angle $\delta_{13}=1.2\pm0.08$, is not particularly small and also comes from the quark mass matrix.  
Indeed, the smallness of CP violation gives us a hint that some dynamical mechanism could be at work to suppress the effects of the $\theta$ term in equation~\eqref{eq:ThetaTerm}. 

\subsection{Axions}

In the year 1977 Peccei and Quinn proposed a mechanism to solve the strong CP problem, often considered to be the most elegant to date~\cite{Peccei:1977hh,Peccei:1977ur}.  The vacuum energy density of QCD depends on $\theta$, i.e., $V_{\rm QCD}=V_{\rm QCD}(\theta)$, and its absolute minimum lies\footnote{Strictly speaking, this would only happen if $\delta_{13}=0$ but CP violation is transmitted by quantum corrections to the QCD sector and the minimum gets shifted from $\theta=0$ by a small amount. See~\cite{Pospelov:2005pr} for a review.} at $\theta=0$~\cite{Vafa:1984xg}, which is CP conserving. 
If $\theta$ is interpreted to be a dynamical field, $V_{\rm QCD}(\theta)$ becomes the potential energy of that field so that the expectation value $\langle \theta\rangle$ will be dynamically driven to zero, explaining the absence of CP violation. 

This mechanism relies on a  global U(1)$_{\mathrm{PQ}}$ symmetry that breaks spontaneously at the PQ scale~$f_a$. A model-independent consequence is that excitations of~$\theta(x)$  around the minimum of the potential represent a new particle, the axion~\cite{Weinberg:1977ma,Wilczek:1977pj}.  The dynamical $\theta(x)$ field needs a kinetic term 
$f_a^2 (\partial_\mu\theta)(\partial^\mu \theta)/2$.
The axion field is the canonically normalized version of~$\theta$, $a(x)=\theta(x) f_a$. 
Values of $f_a\lesssim 10^8$ GeV are excluded experimentally and astrophysically, so 
the axion offers a window to discover physics at ultra-high energies not testable by current accelerator techniques.  

The cancellation of $\langle \theta\rangle$ is dynamical, leading to residual oscillations of $\theta$ around the minimum, which are expected for generic initial conditions. As the age of the universe is finite, these oscillations are quasi-classical field oscillations that could constitute today's cold DM referred to at the realignment mechanism~\cite{Preskill:1982cy,Abbott:1982af,Dine:1982ah}.   

The axion mass is given by
\begin{eqnarray}
\label{axionmass}
m_a&=&5.70(6)(4)\, {\rm \mu eV}\,\left(\frac{10^{12}\rm\,GeV}{f_a}\right), \label{eq:ma}
\end{eqnarray}
where the numbers in brackets denote the uncertainty in the last digit, dominated by the uncertainty in the up-down quark mass ratio (first bracket) and 
higher order effects (second bracket)~\cite{diCortona:2015ldu}. The interaction of the axion with electric fields ${\bf E}$ and magnetic fields ${\bf B}$ is given by the Lagrangian density
\begin{eqnarray}
{\cal L}_{a\gamma} = \frac{\alpha}{2\pi}\,C_{a\gamma} \frac{a}{f_a} {\bf E} \cdot {\bf B}
\quad\quad {\rm with} \quad \quad
C_{a\gamma}=1.92(4) - \frac{{\cal E}}{{\cal N}}\label{eq:cag},
\end{eqnarray}
the fine structure constant $\alpha$, and the model-dependent ratio of the electromagnetic and colour anomalies ${\cal E}/{\cal N}$ of the PQ symmetry. Again the number in brackets refers to the uncertainty in the last digit~\cite{diCortona:2015ldu}.
These numbers are quoted as recently obtained at NLO in chiral perturbation theory~\cite{diCortona:2015ldu}. 
The topological susceptibility entering in equation~(\ref{axionmass}) has also been calculated from first principles exploiting lattice QCD, with a similar uncertainty  (see reference~\cite{nature}). One should however note that $C_{a\gamma}$ might vary by up to two orders of magnitude in somewhat more exotic axion models \cite{DiLuzio:2016sbl,Agrawal:2017cmd}.

\subsection{Landscape and constraints}

\begin{figure}[!]
\centering
\includegraphics[width=12cm]{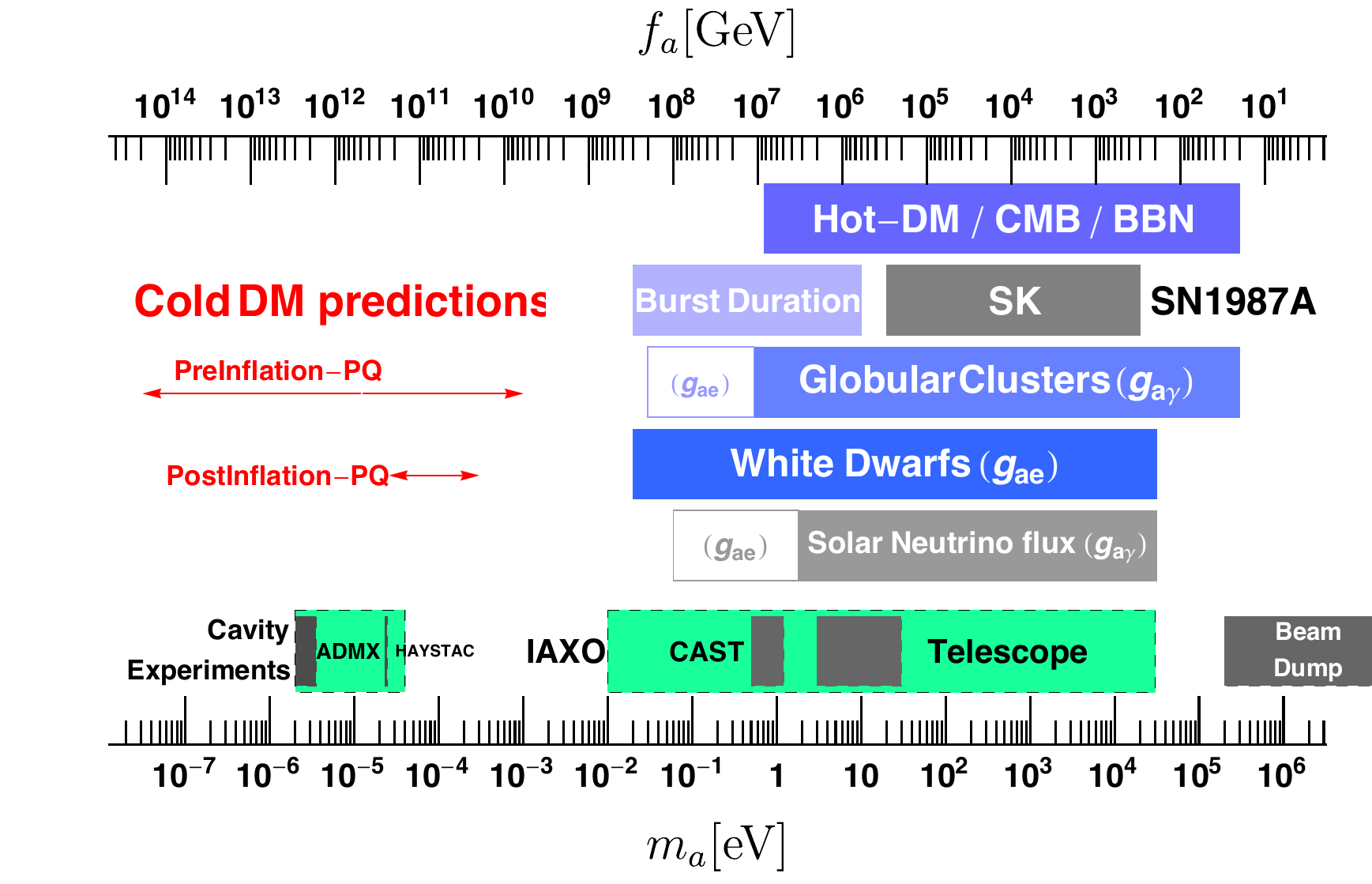}
\caption{Overview of axion masses $m_a$ (equivalently $f_a$) excluded by a variety of astrophysical and cosmological arguments (blue) or laboratory searches (gray) together with experimental prospects (green) and axion cold DM predictions. Pre- and postinflation PQ breaking scenarios are explained in section~\ref{axiondm}.}
\label{qcdbroad}
\end{figure}

The constraints on axion models are usually quoted on a specific coupling, e.g., $C_{a\gamma}$ as a function of $m_a$. A broad picture is shown in figure~\ref{qcdbroad}. A combination of stellar evolution and cosmological arguments together with experimental searches rule out axions with $f_a< 3\times10^8$ GeV corresponding to $m_a>20$~meV.  A significant part of the  axion parameter range is excluded by the impact that axion emission would have in different stellar objects: SN1987A, horizontal branch and red giant stars in globular clusters, white dwarfs and the Sun (see~\cite{raffelt} for a summary and~\cite{Chang:2018rso,Giannotti:2017hny} for updates). Interestingly, some of the observed systems such as white dwarfs, horizontal branch stars in globular clusters and the tip of the red giant branch of the globular cluster M5 show a slight preference for non-standard energy loss and could be hinting at an axion or ALP with $f_a\sim 10^9$ GeV~\cite{Giannotti:2015kwo,Giannotti:2017hny}. 

\begin{figure}[!]
\centering
\includegraphics[width=12cm]{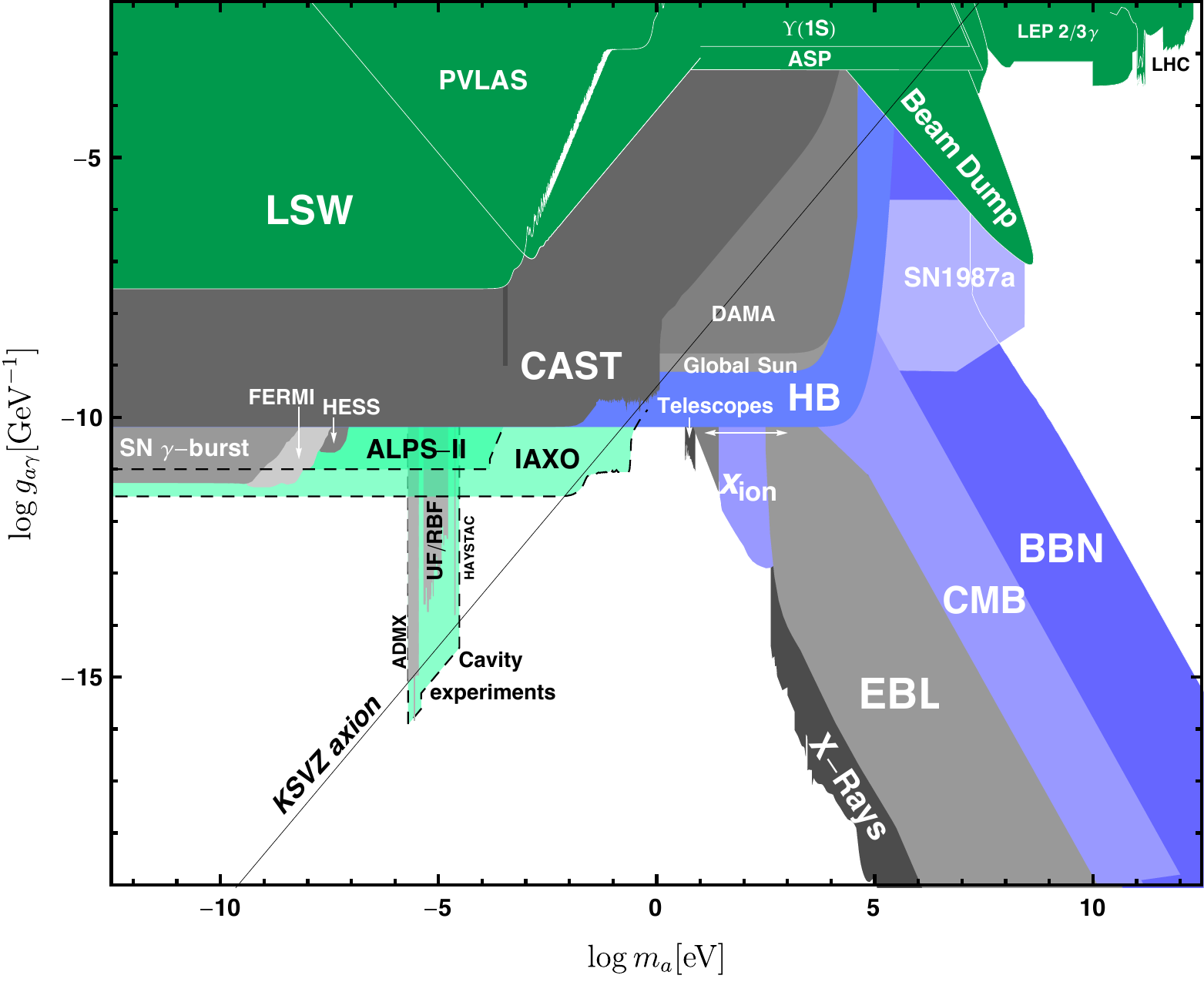}
\caption{Constraints on a generic ALP with mass $m_\mathrm{a}$ and coupling to photons $g_{a\gamma}$. Also shown are the prospects for the reach of IAXO, ALPS-II and the axion DM search experiments ADMX2, ADMX-HF and CAPP.}
\label{ALPbounds}
\end{figure}

The broad picture of experimental limits on any ALP (and including QCD axions such as the KSVZ axion) coupled to photons is shown in the  $m_a$--$g_{a\gamma}$ plane  in figure~\ref{ALPbounds}, where
\begin{equation}
g_{a\gamma} = \frac{\alpha}{2\pi}\frac{C_{a\gamma}}{f_a}. 
\end{equation}
For a recent review on the constraints and prospects for experimental detection of the axion, including this and other couplings, we refer to~\cite{Irastorza:2018dyq}.

\subsection{Axion dark matter \label{axiondm}}

DM axions can be produced in the early universe by at least two processes: in reactions from SM particles in the thermal bath (thermal axions) and by the vacuum realignment mechanism (non-thermal axions)~\cite{Preskill:1982cy,Abbott:1982af,Dine:1982ah}.  
The cold, non-thermal, population is the one that can provide the right amount of cold DM. 
Axion cosmology is reviewed in~\cite{Sikivie:2006ni,Marsh:2015xka}.

In the vacuum realignment mechanism, the axion field starts with certain initial conditions, which then evolves, driven by its potential energy, $V_{\rm QCD}$. The axion DM yield is thus determined by initial conditions and not by thermodynamic processes. 
Two types of axion cosmologies are considered which generally differ in the order of two critical events: cosmic inflation and the PQ symmetry breaking~\cite{1,2}.  

\begin{figure}[!]
\centering
\includegraphics[width=10cm]{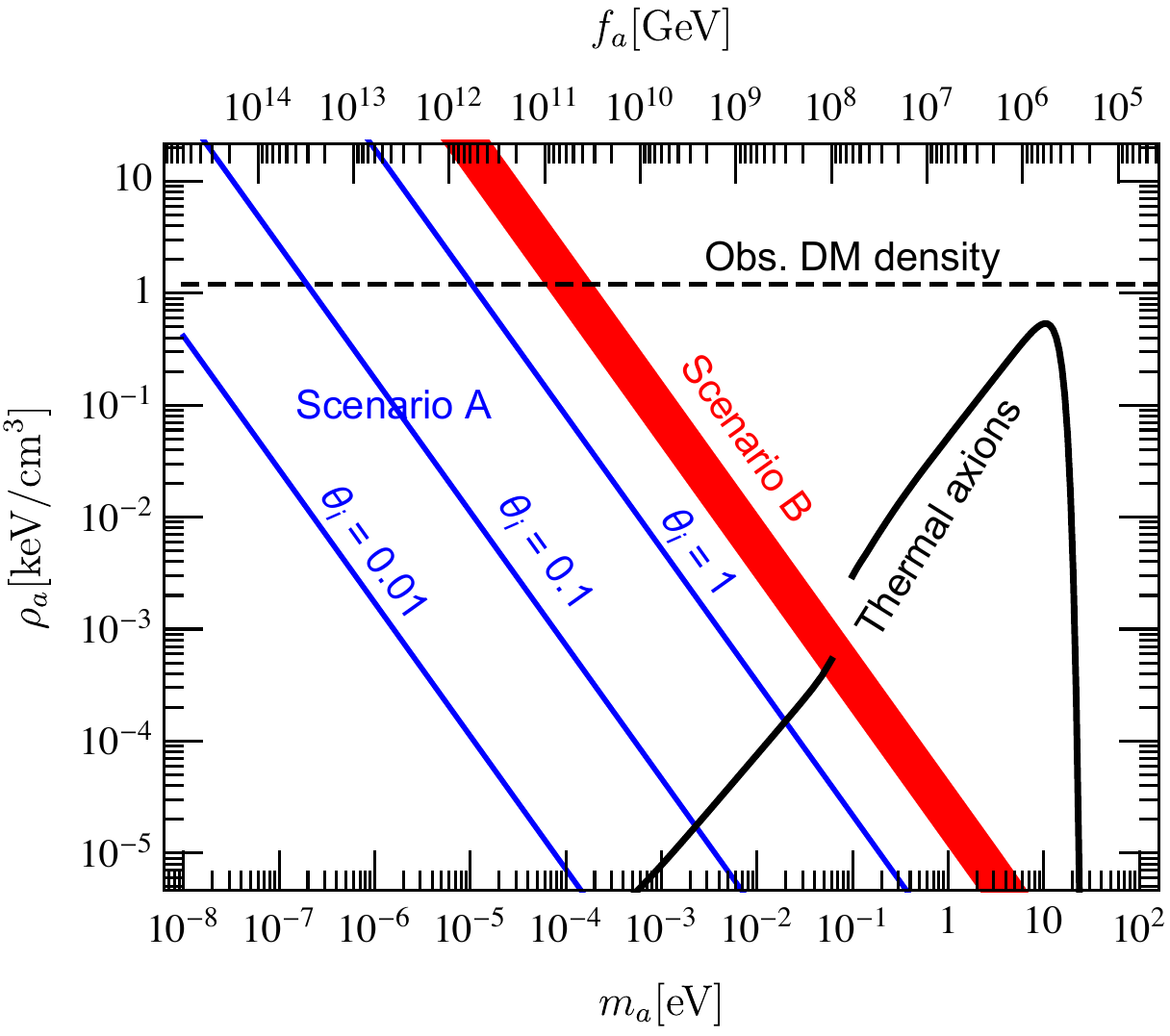}
\caption{Relic axion DM density in the universe today $\rho_a$ as a function of the axion mass $m_a$ (equivalently $f_a$) for several values of the initial misalignment angle $\theta_i$ in scenario A and 
scenario B (${\cal N}=1$) according to~\cite{1,2}. Also shown is the thermal component whose contribution to $\rho_a$ increases towards high mass values.}
\label{fig:axiondm}
\end{figure}

In scenario A inflation happens after the PQ symmetry breaking. One patch is thereby inflated to encompass our observable universe while smoothing~$\theta$ to a single initial value of the misalignment angle~$\theta_i$.
 Accordingly, the initial value of the axion field in our local universe is unique---up to quantum fluctuations---and it is fundamentally  unpredictable from first principles. In this scenario, one can have the complete amount of cold DM in the form of cold axions for any value of $f_a\gtrsim 10^9 \,{\rm GeV}$ assuming a suitable value of the initial universal misalignment angle $\theta_i$, see~\cite{nature}. Several examples for the resulting relic axion density today $\rho_a$ are plotted as blue lines together with the observationally inferred cold DM matter density indicated by the horizontal dashed line in figure~\ref{fig:axiondm}.

In scenario B inflation happens before the PQ symmetry breaking. The axion field keeps a patchy structure from the PQ symmetry breaking until today as the initial conditions of the axion field were essentially random in each causally disconnected patch of the universe. Because of this, one can perform a statistical average to obtain a prediction for the relic axion density today from the realignment mechanism as a function of $f_a$. 
However, scenario B is complicated by the presence of cosmic strings and domain walls, because  the axion field forms a network of global cosmic strings after the PQ symmetry breaking. The ensuing axion domain wall problem is automatically solved when the domain wall number is ${\cal N}=1$ or by a suitable PQ symmetry breaking if ${\cal N}>1$~\cite{Sikivie:1982qv}. 
In processes like string decays some energy converts into long-wavelength axions which will also contribute as DM. 
Assuming ${\cal N}=1$, the scenario B prediction from \cite{2} is shown as a thick red line in figure~\ref{fig:axiondm} (the errors have been slightly enlarged as discussed in \cite{smash}). 
A recent direct calculation with an effective large string-tension technique favours instead a value smaller by a factor of 4, giving the correct abundance for $m_a\simeq 26\, \mu$eV~\cite{Klaer:2017ond}. The recent study about the string network attractor and axion production~\cite{Gorghetto:2018myk} confirms a smaller value than~\cite{2} but highlights that the extrapolations needed are not under control by current simulations so that much larger values are possible. We therefore consider the mass range 
\begin{equation}
26\,\mu{\rm eV}\lesssim m_a \lesssim 1\,{\rm meV}%\mu{\rm meV}
\label{eq:massrangeB}
\end{equation} 
to be best motivated one in scenario~B with $m_a\sim 100~\mu$eV as a typical corresponding value for the DM axion mass. Further discussions are given in~\cite{1,2,Ringwald:2015dsf, Daido:2017wwb, Fleury:2015aca,Fleury:2016xrz,Moore:2016itg,Klaer:2017ond,Gorghetto:2018myk}. 

We will aim to detect the DM axions bound to our galaxy which we assume to provide the full local galactic DM density of $(f_a m_a)^2\theta_0^2/2\sim 300~\mathrm{MeV}/\mathrm{cm}^3$. Their velocity dispersion on Earth is described by the galactic virial velocity $v_a\sim 10^{-3}$. The corresponding de Broglie wavelength is 
$\lambda_{\mathrm{dB}}=2\pi/(m_a v_a)=12.4~\mathrm{m}\,(100~\mu\mathrm{eV}/m_a)(10^{-3}/v_a)$
and thereby of macroscopic size. Indeed, in our axion DM search experiment described below, we expect to probe an axion field that behaves as an (approximately) homogeneous and monochromatic classical oscillating field $\theta\propto \theta_0 \cos(m_a t)$ with $\theta_0\sim 4\times 10^{-19}$ and a frequency of $\nu_a=m_a/(2\pi)$ in the microwave range. 

\section{Foundations of the experimental approach} 
\label{sec:foundations}

The most sensitive experiments to date are based on cavity resonators in strong magnetic fields (Sikivie's haloscopes~\cite{sikivie}) such as ADMX~\cite{admx}, ADMX HF~\cite{admx-hf} or HAYSTAC~\cite{Zhong:2018rsr}. However, these approaches are optimal for $m_a\lesssim 40\,\mu\rm eV$, which has been considered to be a substantial part of the natural range for axion DM in scenario A. If the resonance of the cavity is tuned to the axion mass, the cavity can be understood as a forced oscillator with a large axion-induced
excitation. The length scale of the cavity needs to be approximately $\lambda_a/2$ where $\lambda_a=2\pi/m_a$ is the Compton wavelength given by the axion mass. As the emitted power of the cavity scales with the size of the cavity, this approach is impeded for small wavelengths and therefore small cavity sizes.
For even lower values of $m_a$  nuclear magnetic resonance techniques like CASPER \cite{Budker:2013hfa} or with LC circuits~\cite{Sikivie:2013laa,Kahn:2016aff}, e.g. ABRACADABRA \cite{Ouellet:2018beu} and DM-Radio \cite{Silva-Feaver:2016qhh}, could be effective. 

The mass range favored in scenario~B~\eqref{eq:massrangeB} and $m_a\lesssim 40\,\mu\rm eV$ in particular is not covered by current experiments with a sensitivty sufficiently high to probe QCD axion DM scenarios. In various proposals the cavity concept is extended to this mass range by employing higher mode resonators, such as in ORPHEUS~\cite{Rybka:2014cya}, ORGAN~\cite{McAllister:2017lkb} or RADES~\cite{Melcon:2018dba}. In addition, fifth-force experiments~\cite{Arvanitaki:2014dfa} could search in this region, but would not directly reveal the nature of DM. 

\begin{figure}[!]
\centering
\includegraphics[width=8cm]{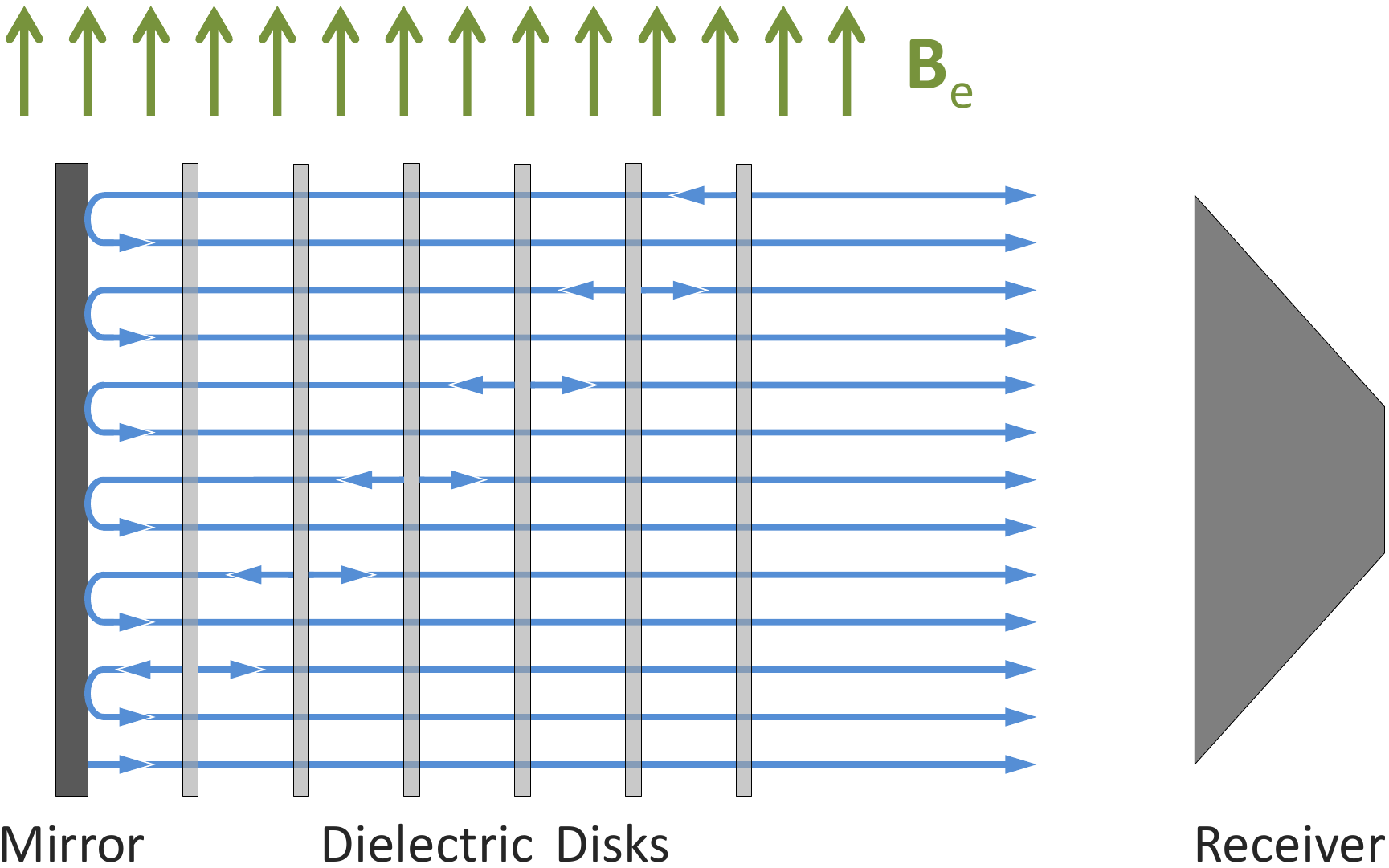}
\caption{A dielectric haloscope consisting of a mirror and several dielectric disks placed in an external magnetic field ${\bf B}_{\rm e}$ and a receiver in the field-free region. A focusing mirror (not shown) could be used to concentrate the emitted power into the receiver. Internal reflections are not shown. Figure taken with permission from~\cite{prl}. 
}
\label{fig:LayeredDielectricHaloscope}
\end{figure}

In~\cite{prl} a new concept to cover this important gap was introduced that is capable of discovering \hbox{$\sim 100\,\mu$eV} mass axions. The main idea of this concept is to exploit constructive interference of electromagnetic radiation emitted at several surfaces as well as resonant enhancement. This is achieved through a series of parallel dielectric disks with a mirror on one side, all within a magnetic field ${\bf B}_{\rm e}$ parallel to the surfaces as shown in figure~\ref{fig:LayeredDielectricHaloscope}---a dielectric haloscope. 

As discussed in \cite{prl} the output power $P$ of the dielectric haloscope per unit area $A$ is 
\begin{eqnarray}
\frac{P}{A} = \beta^2\, \frac{P_{0}}{A}
= 2.2\times 10^{-27}\, \frac{\rm W}{\rm m^2}\,\beta^2
\left (\frac{B_{\rm e}}{10~{\rm T}}\right )^2 C_{a\gamma}^2 \, ,
\label{eq:PoverAdh}
\end{eqnarray}
which implicitly depends linearly on the galactic axion DM density which is here assumed to make up all of the galactic cold DM density; cf.~section~\ref{axiondm}. Moreover, $\beta^2$ is the power boost factor that represents the enhancement of the output power of the dielectric haloscope with respect to the output power $P_0$ of one single magnetized mirror only~\cite{dish}. The value of $\beta=\beta(\nu_a)$ as a function of frequency is calculated by matching 
the axion-induced electric field in each region (dielectric disk or vacuum) with left and right-moving electromagnetic waves of frequency $\nu_a$ as imposed by the continuity of the total electric and magnetic fields parallel to the disk/mirror surface (${\bf E}_{||}$ and ${\bf H}_{||}$, respectively) at the interfaces~\cite{foundations}. The quantum field calculation \cite{Ioannisian:2017srr} of the power agrees with the classical calculation.

The desired enhancement, $\beta^2\gg 1$, comes from two effects, which generally act together but can be differentiated in limiting cases. These effects depend on the optical thickness of each disk $\delta=2\pi \nu d \sqrt{\epsilon} $, where $d$ is the physical thickness, $\epsilon$ the dielectric constant, and $\nu$ the frequency under consideration. This sets the transmission coefficient of a single disk, found to be 
\hbox{${\cal T}=i 2 \sqrt{\epsilon}/[i2\sqrt{\epsilon}\cos\delta+(\epsilon+1)\sin\delta]$}. 
When $\delta=\pi,3\pi,5\pi,...$, the disk is transparent ($|{\cal T}|=1$) and the emission from different disks can be added constructively by placing them at the right distance. 
When $|{\cal T}|< 1$, the spacings can be adjusted to form a series of leaky resonant cavities where $E$-fields are boosted by reflections between the disks. In general, both the simple sum of emitted waves and resonant enhancements are important.

After choosing $\epsilon$ and the thickness of the disks $d$, the distances between disks remain as the only free parameters of the dielectric haloscope,  still leaving considerable control over the frequency response. Different types of configurations are relevant: a configuration with a flat response over a large frequency range $\Delta\nu$ reduces the need for frequent disk repositioning, but suffers from smaller power boost factors $\beta^2$. Configurations with a larger $\beta^2$ over a narrow range $\Delta\nu$ allow to discard statistical fluctuations and do precision axion physics in case of a discovery. 

\begin{figure}[t]
\begin{center}
\includegraphics[width=8.5cm]{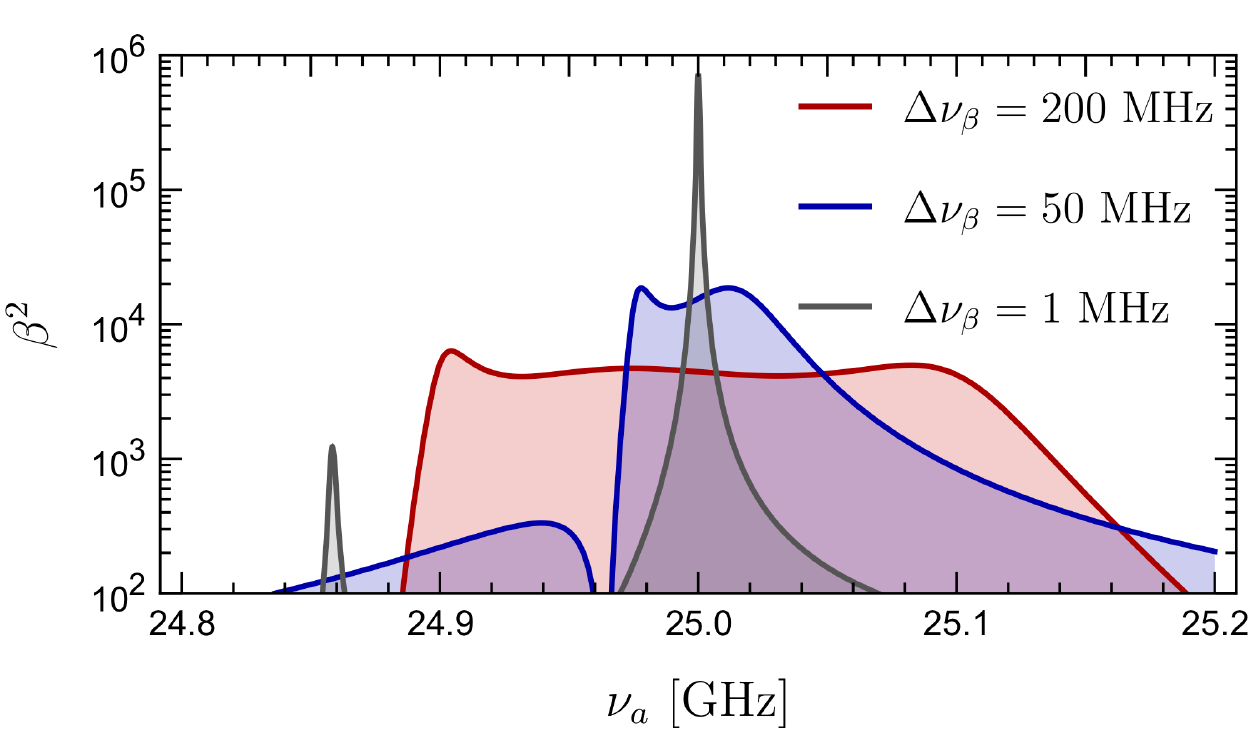}
\caption{Power boost factor $\beta^2(\nu_a)$ for configurations optimized for $\Delta\nu_\beta=200,\, 50\,\,{\rm and}\,\,1~\rm{MHz}$ (red, blue and grey) centered on $25\,{\rm GHz}$ using a mirror and 20~dielectric disks ($d=1\,{\rm mm}$, $\epsilon=25$). 
Figure adapted from~\cite{prl} with permission.
}
\label{fig:bandwidths}
\end{center}
\end{figure}

The behavior of $\beta^2$ can be predicted using the area law: the integral $\int \beta^2 d\nu_a$
is constant for a fixed set of disks, which holds exactly when integrating over \hbox{$0\leq\nu_a \leq \infty$}, and is a good approximation for frequency ranges containing the main peak~\cite{foundations}. This behaviour allows one to  trade width for power and vice versa.
In addition, an increase in the number of disks gives approximately a linear increase in $\int \beta^2 d\nu_a$. 
The area law is illustrated in figure~\ref{fig:bandwidths} which shows $\beta^2(\nu_a)$ for a dielectric haloscope consisting of a mirror and 20 disks (\hbox{$d=1~{\rm mm}$}, \hbox{$\epsilon=25$}). Spacings have been selected to maximize the power boost factor $\beta^2$ for three ranges of $\Delta\nu$ with  \hbox{$\Delta\nu_\beta=1,\,50\,\,\rm{and}\,\,200~{\rm MHz}$} each equally centered on $25~{\rm GHz}$. 

\section{Proposed experimental setup for the search of dark matter axions}
\subsection{Design sensitivity and constraints from technology}
The goal is to build a dielectric haloscope, based on the experimental concept described in section \ref{sec:foundations}, that is sensitive to axion DM  in the mass range \MassRange. The corresponding frequency range to be covered is roughly \FrequencyRange. 
The feasibility of achieving this goal is discussed in the following sections based on equation~(\ref{eq:PoverAdh}) and considering the constraints imposed by available technologies and materials.  

Current state of the art electromagnetic receiver systems are able to detect  signal powers of roughly \BenchmarkPower for a few days measurement time and for frequencies below ~40~GHz. Such receiver systems are also used in radio astronomy applications. They have noise temperatures $T_{\rm rec}$ of a few~K. More details about the receiver are discussed in section \ref{sec:receiver}.

The magnetic field ${\bf B}_{\rm e}$ needs to be parallel to the disk surface as introduced in section~\ref{sec:foundations}. This requires ideally a dipole magnet that encloses the entire booster setup. To obtain a detectable power emission, a minimum value for the figure of merit  $B^2A$ is \MagnetBsquareA, where $B=B_{\rm e}$ here and below. The magnet will be discussed in more detail in section \ref{sec:magnet}. 

The  disks need to have high dielectric constants $\epsilon$, and small dielectric losses $\tan \delta$. They need to be mechanically stable such that  disks with large surfaces and a few millimeter thickness can be manufactured. Several materials are considered for this purpose, for example LaAlO$_3$, with $\epsilon \approx \EpsiolonLaAl$ and $\tan \delta \approx \TanDeltaLaAl$ at low temperatures \cite{lalo3_a,lalo3_b}. It seems realistic to manufacture tiled LaAlO$_3$ disks of significant size, which is discussed in more detail in section \ref{sec:booster}.

With these technological constraints, it follows from equation \eqref{eq:PoverAdh} that the power boost factor $\beta^2$ needs to exceed a value of $\sim 10^4$ to make an axion signal detectable. The expected sensitivity and the measurement strategy are discussed in more detail in the sections below.

\subsection{The receiver \label{sec:receiver}}

\begin{figure}[t]
\centerline{
\includegraphics[width=0.4\textwidth]{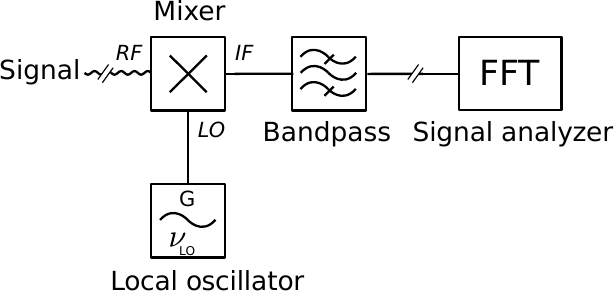}
}
\caption{\label{fig:heterodyne}
Basic block diagram for heterodyne detection. By mixing the signal at frequency $\nu$ with a local oscillator signal at frequency $\nu_{\rm LO}$, the signal is shifted to $\nu + \nu_{\rm LO}$ and $\nu - \nu_{\rm LO}$. One of them is selected with a band-pass and successively detected.
}
\end{figure}

The proposed principle of microwave detection for the frequency regime below $\sim$\,\SI{40}{\giga\hertz} is based on heterodyne mixing of a preamplified signal \cite{heterodyne}. This is sketched in figure \ref{fig:heterodyne}. After the first low noise preamplifier stage (HEMT) and pre-filtering, the signal (RF) is  shifted to intermediate frequencies (IF) by mixing a carrier frequency into the signal. At the intermediate frequencies the signal is further amplified and filtered. Finally, the signal is at a frequency that is accessible to digital 16 bit samplers, which have internal FPGAs that provide real time fast Fourier transform calculation with subsequent integration and storage of the signal.

As mentioned above, state of the art detector technology requires different systems
for the frequency ranges \SIrange{10}{40}{\giga\hertz} and above 40~GHz.
For the lower frequencies, HEMT detectors~\cite{hemt}, as widely used in the radio astronomy community, can be utilized. For the baseline design we propose  HEMT receivers from Low Noise Factory\footnote{www.lownoisefactory.com}
for the lower frequency range. This frequency range is the initial focus of the experiment motivated by the mass region that is predicted for the DM axion, as discussed in section \ref{axiondm}. For the high frequency range new detectors 
working at the quantum noise limit or below still have to be identified and developed.

The sensitivity of the receiver can then be calculated using Dicke's radiometer equation \cite{dicke,dicke2},
\begin{equation}
\label{eq:sens}
\frac{S}{N}=\frac{P_{\rm sig}}{k_B T_{\rm sys}} \sqrt{\frac{t_{\rm scan}}{\Delta\nu}}
\end{equation}
where $k_B$ is the Boltzmann constant. $T_{\rm sys}$ is the total system noise temperature which consists of the receiver noise temperature $T_{\rm rec}$ and the additional noise from the booster and its surroundings $T_{\rm booster}$, such that $T_{\rm sys} = T_{\rm rec} + T_{\rm booster}$. $P_{\rm sig}$ is the expected photon signal power \eqref{eq:PoverAdh} corrected by the experimental efficiency, $t_{\rm scan}$ is the scanning time for an individual measurement of a given bandwidth, and $\Delta\nu=10^{-6}\nu$ is the predicted axion line width at given frequency $\nu$.

Given the state of the art receiver noise temperature  of $T_{\rm rec}\approx \TSysReceiver$ the noise  of the remaining system, in particular the booster, should not exceed a few K. For the  discussion in the following sections a value for the booster noise  of $T_{\rm booster} \approx \TSysBooster$ will be used.	 The noise  of the booster is calculated from the actual physical temperature multiplied by its emissivity, i.e., its effectiveness in emitting thermal radiation. The emissivity equals the absorption coefficient. A perfect mirror would have no absorption and therefore a noise temperature of zero. Realistic systems with lossy disks will have finite emissivity and will therefore need cooling. Furthermore, the antenna and supporting structures thermally radiate. This can be suppressed by cooling the whole setup. These requirements can  only be achieved with the booster being enclosed in a cryostat. With these boundary conditions the benchmark power of \BenchmarkPower can be detected by the receiver within a few days measurement time. 

\subsection{The magnet\label{sec:magnet}}

According to equation~(\ref{eq:PoverAdh}), the emitted power is proportional to the square of the
magnetic field component $B$ parallel to the surface and the area $A$ of the surface.
Together with the necessity to collect the generated power by antennas facing the surfaces of the disks,
this implies that a dipole field  is  preferred. When designing a magnet for the haloscope, the quantity $B^2A$ is
 to be maximized. At the same time, the maximum length and width of the setup imposed by the coherence requirement,
signal attenuation in the disks, and mechanical constraints need to be considered.

Taking the
discussion in section \ref{sec:booster} into account, a
dipole magnet suitable for the experiment should reach a $B^2A$
 value of \MagnetBsquareA. This could be realized with
a magnetic field strength of  \MagneticField, with a bore of
\MagnetBore allowing to host disks of similar size. As discussed in section \ref{sec:booster} this field
should extend  over a length of up to \MagnetLength.

Two independent conceptual design studies are presently being finished.
They are performed in the framework of an innovation partnership
\cite{bibinnovationpartner}.
%with Bilfinger Noell\footnote{Bilfinger Noell Magnettechnik, Alfred-Nobel-Str. 20 97080 W\"urzburg, Germany} and the magnet group at CEA-Irfu Saclay\footnote{DRF/IRFU/DACM -- CEA Saclay, Bat. 123, 91191 Gif sur Yvette, France}.

Both innovation partners have investigated several different dipole-magnet concepts:
cosine-theta \cite{cosine}, canted cosine-theta \cite{canted}, racetrack \cite{racetrack} and block designs \cite{block}.
They independently came to the conclusion
that it is technologically feasible to produce a dipole magnet compatible with the
required $B^2A$ value of \MagnetBsquareA and a field homogeneity of 5\%
within the geometrical boundary conditions set by the experiment
and the infrastructure at the planned experimental site at DESY, Hamburg.
Such a magnet would be built according to the block design
using NbTi superconductor at 1.9~K. In order to respect the maximum peak field
inside the coils consistent with NbTi superconductor,
the magnetic field would be 9\,T, while the disk diameter would be 1.25\,m.

As a first step a demonstrator coil for verification of
the feasibility of the proposed conceptual design will be built.
First estimates of the  time schedule indicate the possibility for
a delivery of the full scale magnet to the experimental site around 2025.

\subsection{The booster \label{sec:booster}}

The requirements for the booster follow from the design sensitivity and the constraints from receiver and magnet technologies which were discussed in the sections above. The signal power of the booster needs to be of the order of $\gtrsim\,$\BenchmarkPower to be detectable by state of the art radiometers. According to equation~(\ref{eq:PoverAdh}), the factor $\beta^2$ needs to exceed four orders of magnitude with a $B^2A$ value of \MagnetBsquareA, for example with a disk size of \MagnetBore and magnetic field of \MagneticField.

\begin{figure}
\centerline{
\includegraphics[width=0.75\textwidth]{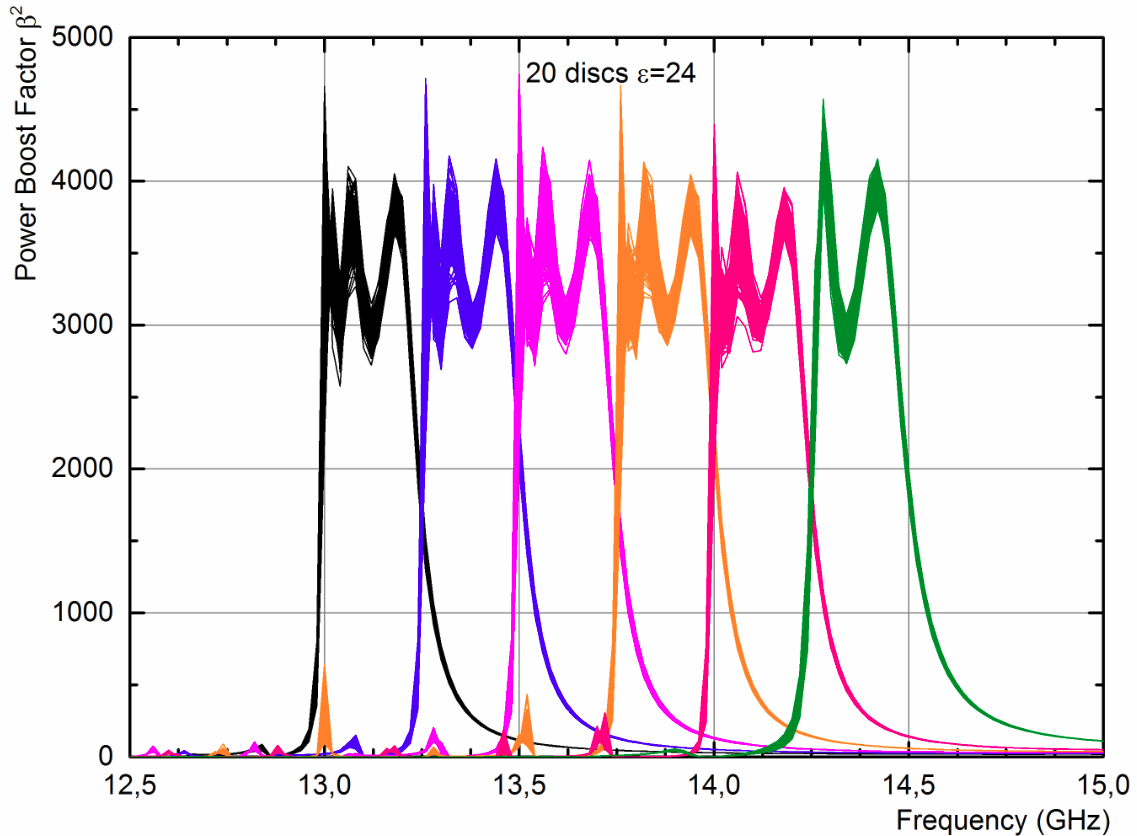}
}
\caption{Calculations of the power boost factor $\beta^2$ with 20 LaAlO$_3$ disks ($\epsilon$=\EpsiolonLaAl) for six different disk spacing configurations. For each configuration $\beta^2>2.5\times 10^3$ is achieved over a range of  $\Delta\nu_\beta\sim 250$~MHz. The 250 simulations performed with random Gaussian variations of the disks spacings with a precision of $\sigma$=15\,$\mu$m are plotted on top of each other. 
}\label{fig:boost}
\end{figure}

Following the calculations outlined in section \ref{sec:foundations}, the power boost factor $\beta^2$ is determined using an idealized scenario with planar disks and without diffraction.  Figure~\ref{fig:boost} shows the power boost factor $\beta^2$ as a function of frequency resulting from these calculations using 20 disks made from LaAlO$_3$ ($\epsilon=24)$. Six different configurations of disk spacings have been used. For each configuration the disk spacings are chosen such that the boost factor exceeds $\beta^2 > 2.5\times 10^3$ within a frequency range of $\Delta\nu_\beta \sim$\,250\,MHz. 
For each configuration the calculation was repeated 250 times  with Gaussian variations of the disks spacings with a precision of $\sigma=15\,\mu$m. This indicates that position uncertainties of 15\,$\mu$m are well acceptable in this frequency range, changing $\beta^2$ typically less than $\sim\SI{5}{\percent}$.

The frequency at which sizable $\beta^2$ values occur in the range $\Delta\nu_\beta$
can be seamlessly shifted by changing the spacings between the disks. The area law then implies that for a narrower $\Delta\nu_\beta$ setting a higher power boost $\beta^2$ can be achieved.
Further idealized simulations showed that the power boost $\beta^2$ can easily exceed \BenchmarkPowerBoost for a setup with 80 disks~\cite{Millar:2017eoc}.  Given the axion mass range to be covered, the distances between the disks range from $\sim$~\SI{1.5}{\milli\meter} at \SI{100}{\giga\hertz} to $\sim$~\SI{20}{\milli\meter} at \SI{10}{\giga\hertz}. The number of 80 disks would necessitate a length of the system of up to 200~cm (low frequency setup).

The booster could consist of movable disks connected to precision rails on which they can be positioned by precision motors via pistons. While the precision of the pistons of motors themselves can be easily controlled to the sub $\mu$m level, the mechanical transmission from motors to disks in a high magnetic field and cryogenic environment as well as gravity can lead to sizable uncertainties in the exact disk positioning. 
Technology to ensure in-situ adjustable disk spacing with high enough precision in the experimental 
surrounding with \MagneticField magnetic field and cryogenic ambient temperature is currently being investigated and developed.

The material used for the dielectric disks has to fulfill the following criteria:
\begin{itemize}
\item High dielectric constant $\epsilon \gtrsim$ \num{10}: 
As discussed in \cite{foundations} the output power can be increased with a higher dielectric constant $\epsilon$, as the discontinuity of the axion-induced field on a disk surface increases.
In addition, a higher $\epsilon$ can enhance the boost factor by making the system more resonant.
Moreover, the disks are more difficult to manufacture for high $\epsilon$ as they need to be thinner for a given optical thickness $\delta$ and their placement more precise.
The optimal $\epsilon$ is therefore a trade off between various effects. 
\item Low dielectric loss tan\,$\delta \lesssim 10^{-5}$: Similar calculations as in section \ref{sec:foundations} and \cite{foundations}  have shown that up to 50\% of the total output power in a 80 disk booster is lost for $\tan{\delta} = 10^{-6}$ in the most resonant configurations. For more broadband configurations, higher losses with up to $\tan\,\delta \sim 10^{-5}$ are acceptable. 
\item Mechanically stable.
\item Appropriate cryogenic properties down to \TActualBoosterTemp.
\item Affordable.
\end{itemize}
Presently lanthanum aluminate (LaAlO$_3$) is envisioned as the baseline candidate for its high permittivity ($\epsilon\approx$~\EpsiolonLaAl) and small loss ($\tan\delta\approx$~\TanDeltaLaAl) also at low temperatures \cite{lalo3_a,lalo3_b}. An alternative material could be sapphire (Al$_2$O$_3$) \cite{al2o3_a,al2o3_b} which has a permittivity of $\epsilon\approx$~\EpsiolonSaphire and a loss angle $\tan\delta\approx$~\TanDeltaSaphire. Other materials are under investigation. 

The technology to produce sufficiently large disks with high enough precision (surface roughness of $\sim \mu$m) needs to be developed. The concept of disk tiling is currently under investigation. Tiled disks are made from several smaller pieces of dielectric material that are glued or connected otherwise to form a single, stable large disk. Preliminary results from 3D simulations show that the emitted power of a tiled disk with gaps of $\lambda_a / 10$ is not significantly reduced compared to the one emitted by a monolithic disk. Indeed, the emitted power is only reduced according to the smaller area of the tiled disk within the uncertainties of the simulation. 
Also first transmission measurements with a tiled ceramic disk showed no measurable effect.

An important task is to suppress contributions to the noise temperature $T_{\rm sys}$ from the booster and its surrounding.
As discussed in section \ref{sec:receiver}, the added system noise needs to be less than $\sim$\,\TSysBooster. The main noise component is expected from thermal radiation of the disks with the support system, the walls surrounding the booster and the antenna.  Therefore the booster needs to be enclosed in a cryogenic environment that allows to operate the booster at a temperature of \TActualBoosterTemp.

\begin{figure}
\centerline{
\includegraphics[width=1\textwidth]{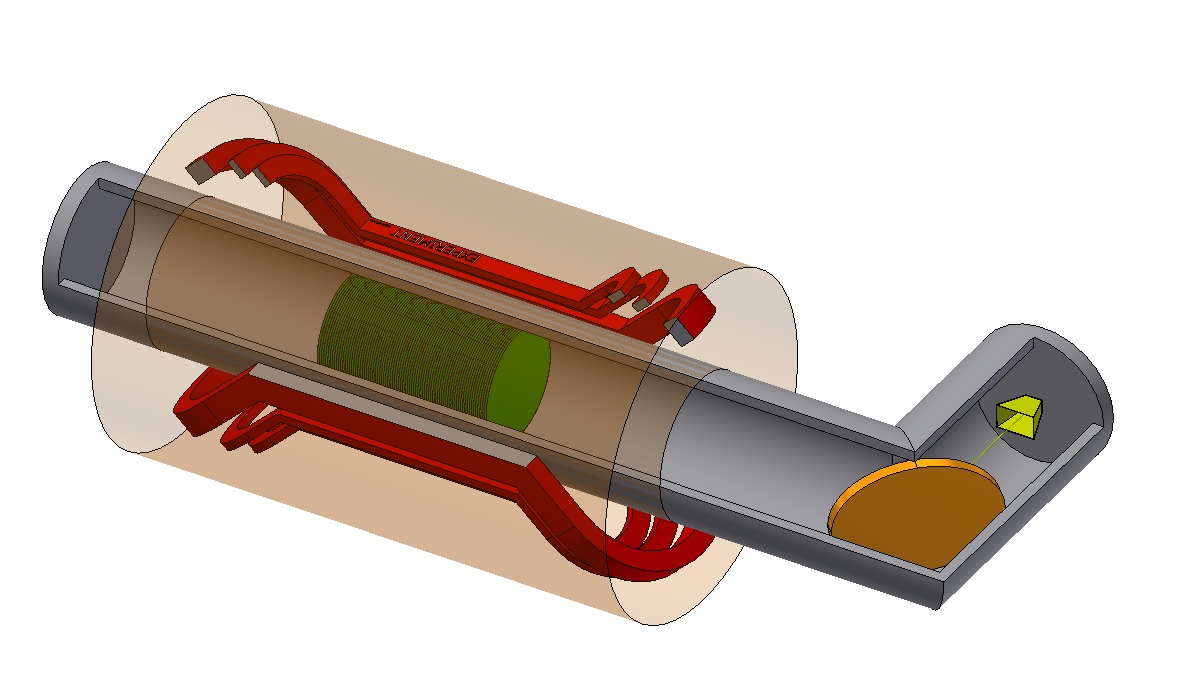}
}
\caption{Preliminary baseline design of the MADMAX approach. The experiment can be divided into three parts: (1) magnet (red racetracks), (2) booster -- consisting of the mirror (copper disk at the far left), the 80 dielectric disks (green) and the system to adjust disk spacing (not shown) -- (3) the receiver -- consisting of the horn antenna (yellow) and the cold preamplifier inside a separated cryostat. The  focusing mirror is shown as an orange disk at the right. The figure is not to scale.
}\label{fig:baseline_design}
\end{figure}

A baseline design with 80 disks is assumed for the sensitivity study in section~\ref{sec:strategy}.  A schematic of the proposed MADMAX approach is shown in figure~\ref{fig:baseline_design}.
The experimental setup can be divided into the three main components: 
\begin{itemize}
\item The dipole magnet with a $B^2A$ value of \MagnetBsquareA over a length of 2~m, 
\item the booster, consisting of a mirror ($\epsilon = \infty$) at the far end and the $\sim$\,80 dielectric disks that can be positioned within a few $\mu$m precision by motors,
\item the receiver, including the focusing mirror and the antenna, which is used for detection of the emitted power.
\end{itemize}

As discussed in section \ref{sec:receiver} the initial focus of the experiment is in the frequency range \SIrange{10}{40}{\giga\hertz}. The extension to higher frequencies will be done at a later stage. For higher frequencies diffraction effects are expected to become less pronounced but the mechanical precision requirements on the booster setup will become more demanding. 

\subsection{Expected sensitivity and measurement strategy \label{sec:strategy}}

The  sensitivity of the proposed setup is calculated from the baseline design discussed in the previous sections. 
This estimate assumes that a power boost factor of $\beta^2$\,\BenchmarkPowerBoost over a bandwidth of $\Delta\nu_{\beta}\sim$\,\BenchmarkBoostWidth~can be achieved with \DiskNumber~LaAlO3 disks with a surface area of \MagnetBore~in a \MagneticField magnetic field.
For the receiver a detection efficiency of \DetectionEfficiency and a measurement time of \BroadbandMeasurmentTime~, system noise temperature of~\TSysTotal~and minimum signal-to-noise ratio of $S/N = 4$ for each frequency band are assumed.

\begin{figure}
\centerline{
\includegraphics[width=0.9\textwidth]{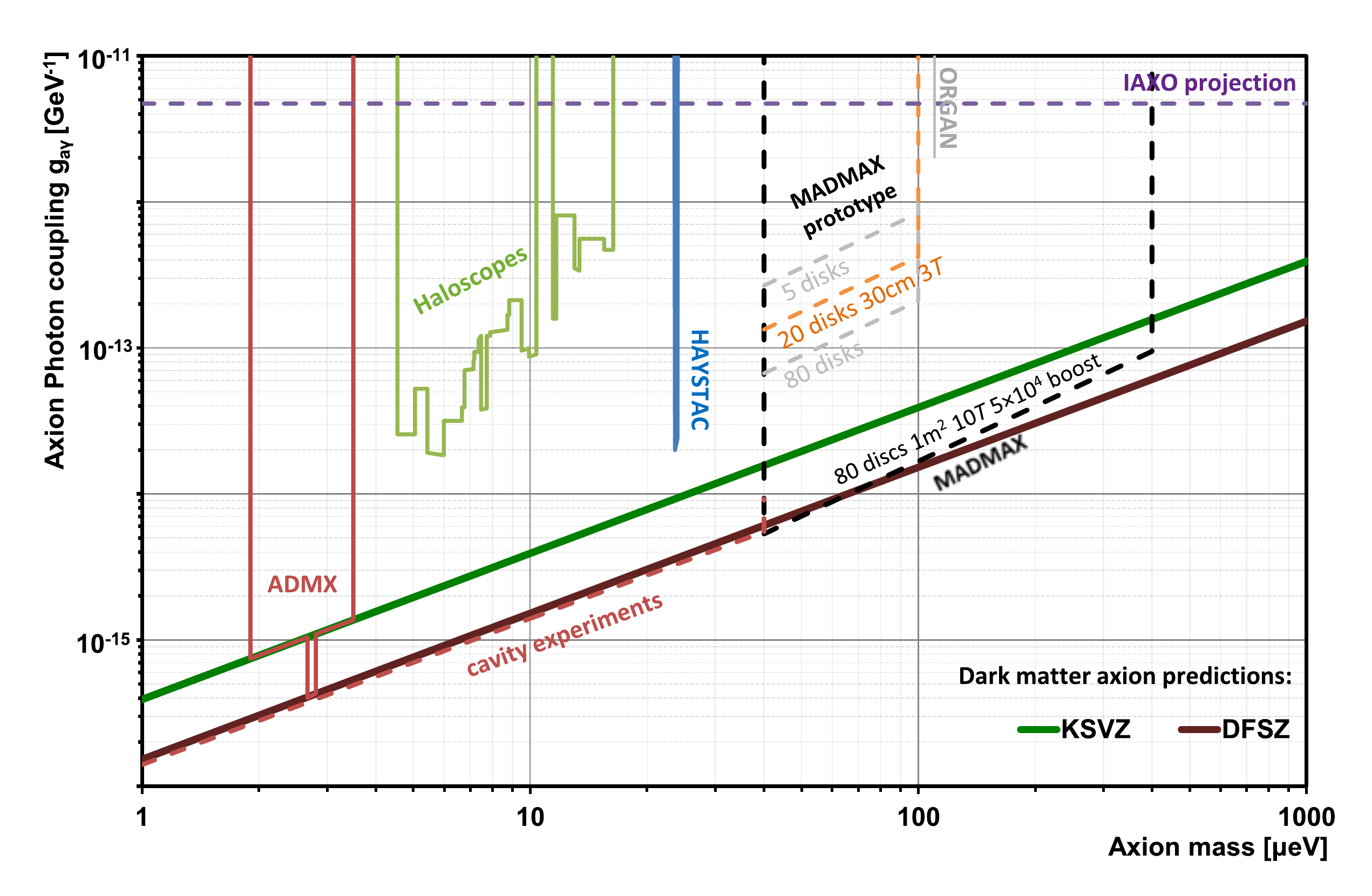}
}
\caption{\label{fig:sensitivity}
Projection of sensitivities for DM axions and ALPs on the axion-photon coupling $g_{a\gamma}$ as a function of the axion mass. The lines denoted by DFSZ and KSVZ show representative $g_{a\gamma}$ values associated with the corresponding two most popular classes of QCD axion models.
%For all limit setting sensitivities an axion line width of 10$^{-6}$, 8\,K total noise temperature of the benchmark system and 4\,$\sigma$ separation from background have been assumed. 
For the MADMAX projection, a setup with 80 dielectric disks each with an area of \MagnetBore and a magnetic field of  \MagneticField parallel to the disk surfaces has been assumed.
Also shown are sensitivity projections for prototype setups with 5 and 20 disks both with 30~cm disk diameter in a 3~T magnetic field.
In addition, an approximate projection of the reach of future cavity experiments is shown.
The projected sensitivities are compared to existing limits from ADMX \cite{admx,Du:2018uak}, other haloscope experiments \cite{RingwaldPDG}, from HAYSTAC \cite{admx-hf} and ORGAN \cite{McAllister:2017lkb}.
Also the IAXO \cite{iaxo} sensitivity for solar axions and ALPs is indicated.}
\end{figure}

\begin{figure}
\centerline{
\includegraphics[width=0.9\textwidth]{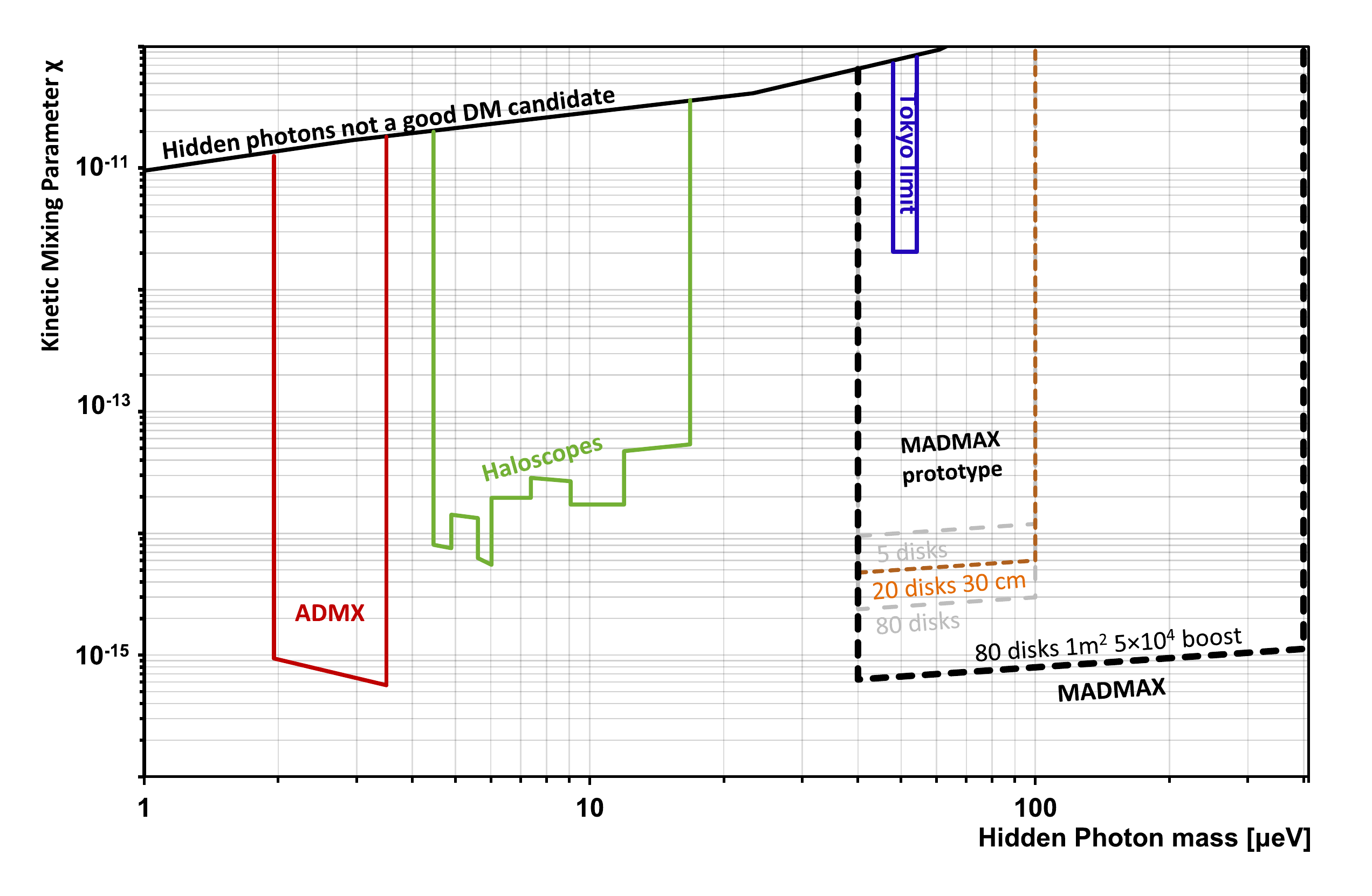}
}
\vspace{-0.5em}
\caption{\label{fig:sensitivity_hp}
Projection of sensitivities for hidden photons on the kinetic mixing parameter $\chi$ as a function of the hidden photon mass. 
%For all limit setting sensitivities a hidden photon line width of 10$^{-6}$, 8\,K total noise temperature of the  system and 4\,$\sigma$ separation from background have been assumed. 
For the MADMAX projection, a setup using 80 lanthanum aluminate disks each with an \MagnetBore area have been assumed.
Also shown are sensitivity projections for prototype setups with 5 and 20 disks with 30~cm disk diameter.
The projected sensitivities are compared to existing limits from ADMX and other haloscope experiments for hidden photon searches from~\cite{hidden_photons}, from ADMX-HF~\cite{admx-hf} and from the latest Tokyo results~\cite{tokyo_limits}.
The area above the black line represents the parameter space where hidden photons are not a good DM candidate~\cite{hidden_photons}.
}
\end{figure}

The resulting sensitivities obtainable with the MADMAX approach for DM axions and hidden photons with a mass in the range \MassRange using the above assumptions and the Dicke equation~\eqref{eq:sens} are shown in figures~\ref{fig:sensitivity}~and~\ref{fig:sensitivity_hp}, respectively. 
A significant part of the parameter space predicted for DM axions in the post inflationary PQ scenario could be probed. For hidden photons, the kinetic mixing angle $\chi$ is analogous to $g_{a\gamma}$ and the search does not require a magnetic field~\cite{Jaeckel:2013ija,dish}. Also a  parameter space consistent with hidden photons as DM could be probed down to $\chi \lesssim 10^{-15}$.

The measurement strategy depends on %the sensitivity of the receiver system, i.e. the
the required measurement time to detect a \BenchmarkPower signal with the desired   significance and the time needed to readjust the disk spacings to change the frequency band of the power boost factor $\beta^2$.
%The characteristic properties of the boost factor curve  is  the frequency range $\Delta\nu_\beta$ in which the power boost is enhanced, and the $\beta^2$ value itself. %An optimal scan will require that the measurement time is approximately the same as the time needed to readjust the disk spacings.
The Dicke equation \eqref{eq:sens} suggests that the measurement time for a scan over a fixed frequency range is inversely proportional to the square of the signal power $P_{\rm sig}$, where $P_{\rm sig}$ itself is proportional to $\beta^2$. The measurement time is further proportional to the number of individual scans per GHz, given by $1 \rm  \ GHz / \Delta \nu_\beta$, such that the total measurement time is inversely proportional to $\beta^4\Delta \nu_\beta$.
Since the area law suggests $\beta^2 \Delta \nu_\beta \sim {\rm const}$, it is in principle favorable to make $\beta^2$ big and $\Delta \nu_\beta$ small, until the measurement times become comparable to the readjustment times for the disks. Increasing the boost factor further decreases $\Delta \nu_\beta$ and thus requires more readjustments, which in turn increase the total scan time again.

The readjustment time, during which no data is taken, is conservatively estimated to be around a day. 
The measurement time further depends on the system noise which is assumed to be $T_{\rm sys} = $ \TSysTotal~and the detection efficiency, for which we assume $\sim\DetectionEfficiency$.
Note that the numbers stated in the above considerations are preliminary estimates with respective uncertainties.
Under the stated assumptions one finds that the mass range between 40 and 120~$\mu$eV could be scanned within {$\sim5$\,years} \cite{foundations}, showing the experimental feasibility of this approach.
The mass range above 120~$\mu$eV requires additional R\&D especially for the detection technique and could be covered with  detectors working at or below the quantum limit for the required frequency range.

\section{Proof of principle measurements with first test setup}
First proof of principle systems for the booster and the receiver have been assembled and tested.
A receiver system based on a HEMT preamplifier and heterodyne mixing as described in section~\ref{sec:receiver} has been set up.
Additionally, a first proof of principle booster with up to 5 sapphire disks has been built and the electromagnetic response was tested for different cases of equidistant disk positions.

The setups were used for first proof of principle measurements, and are described in more detail in the following subsections. These measurements indicate that the assumptions on 10\,--30\,GHz receiver sensitivity and disk placement precision, necessary to estimate the sensitivity of the MADMAX approach, are realistic.

\subsection{The proof of principle detection system \label{sec:receivertest}}
\begin{figure}
\centerline{
\includegraphics[width=1.0\textwidth]{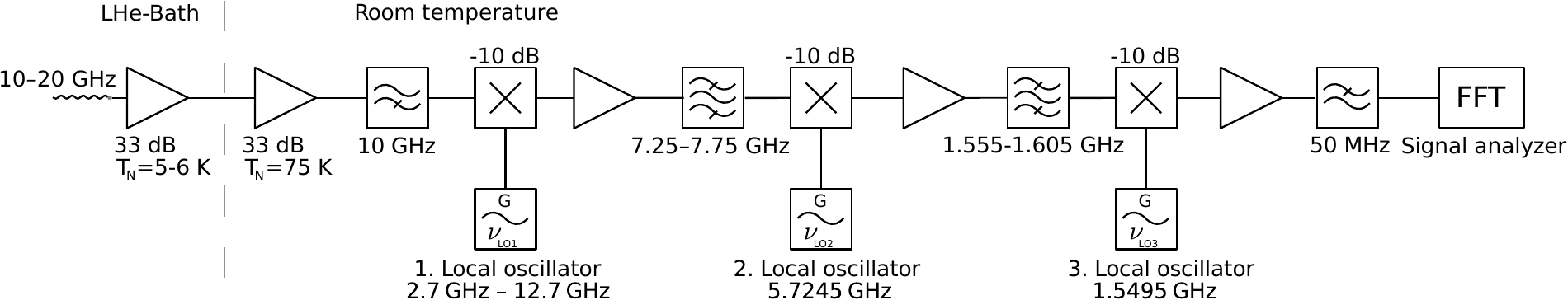}
}
\caption{Receiver chain of the presented proof of principle receiver. 
}\label{fig:proofofprinciple_receiver_chain}
\end{figure}

The detection system of the ``proof of principle setup'' consists of a 3-stage heterodyne receiver with subsequent Fast-Fourier signal analysis (figure \ref{fig:proofofprinciple_receiver_chain}). The high sensitivity of the receiver is achieved by an InP-HEMT operating at an ambient temperature of \SI{4}{\kelvin}. The amplifier has a noise temperature of \SI{5}-\SI{6}{\kelvin} and a gain of \SI{33}{\dB}, which is sufficient to determine the noise performance of the receiver. Afterwards the frequency down-conversion is done at room temperature to a center frequency of \SI{26}{\mega \hertz} with a total bandwidth of approximately \SI{50}{\mega \hertz}. Data acquisition happens using four time-shifted digital 16 bit samplers with a sampling rate of $200\times 10^6$/s and internal FPGAs for real time FFT calculation and subsequent averaging.  This method allows the reduction of the system dead-time from 75\% to less than 1\%. The bin-width of the Fourier transform is \SI{2.048}{\kilo \hertz}.

\begin{figure}
\centerline{
\includegraphics[width=0.6\textwidth]{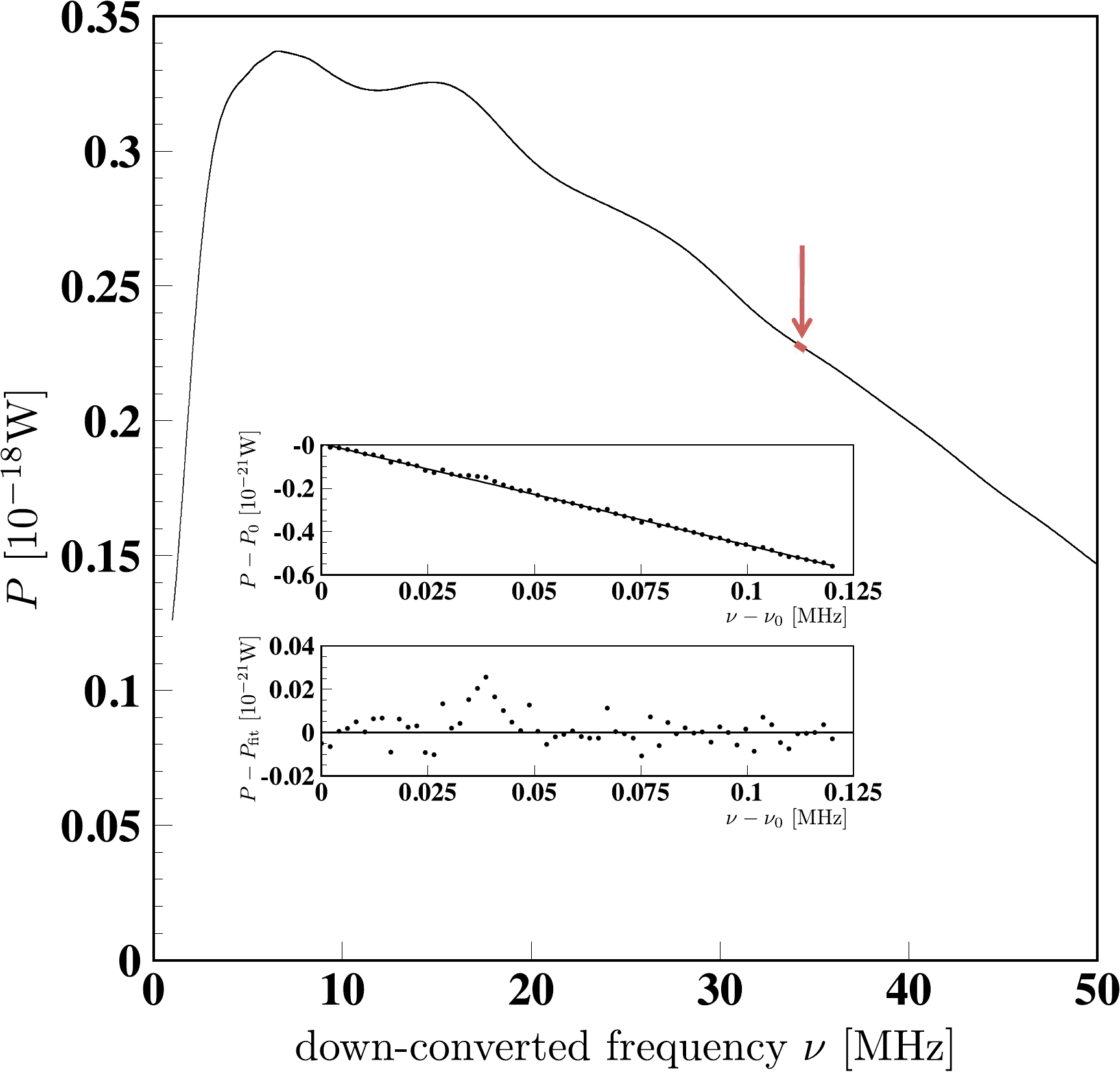}
}
\caption{Test signal at \SI{18.4}{\giga \hertz} and a power of $\sim\,1.2\times10^{-22}$~W recorded with the presented receiver setup.
The region of the signal is shown in the inset together with the background fit (top inset) and the residuals from the background fit (lower inset). The test signal is detected with a significance of $\gtrsim$\,4.8\,$\sigma$. The arrow indicates the location of the inset. Adapted from \cite{receiver_statistics_paper}.
}\label{fig:proofofprinciple_receiver_powerspectrum_adapted}
\end{figure}

First measurements with the receiver have been performed. A ``fake-axion'' signal with a level of $\sim\,1.2\times10^{-22}$~W and a bandwidth of about \SI{10}{\kilo \hertz} was injected. The \SI{4}{\kelvin} background temperature of the signal injection was determined by the liquid helium environment. This fake axion signal could be detected in a cross-correlation within two days measurement time with $\gtrsim$\,4.8\,$\sigma$ separation from background. The power spectrum is shown in figure \ref{fig:proofofprinciple_receiver_powerspectrum_adapted}. The receiver thus meets the first expectations. Further improvements to reduce the system noise temperature are ongoing.

\subsection{The proof of principle booster \label{sec:boostertest}}
\begin{figure}
\centerline{
\includegraphics[width=0.5\textwidth]{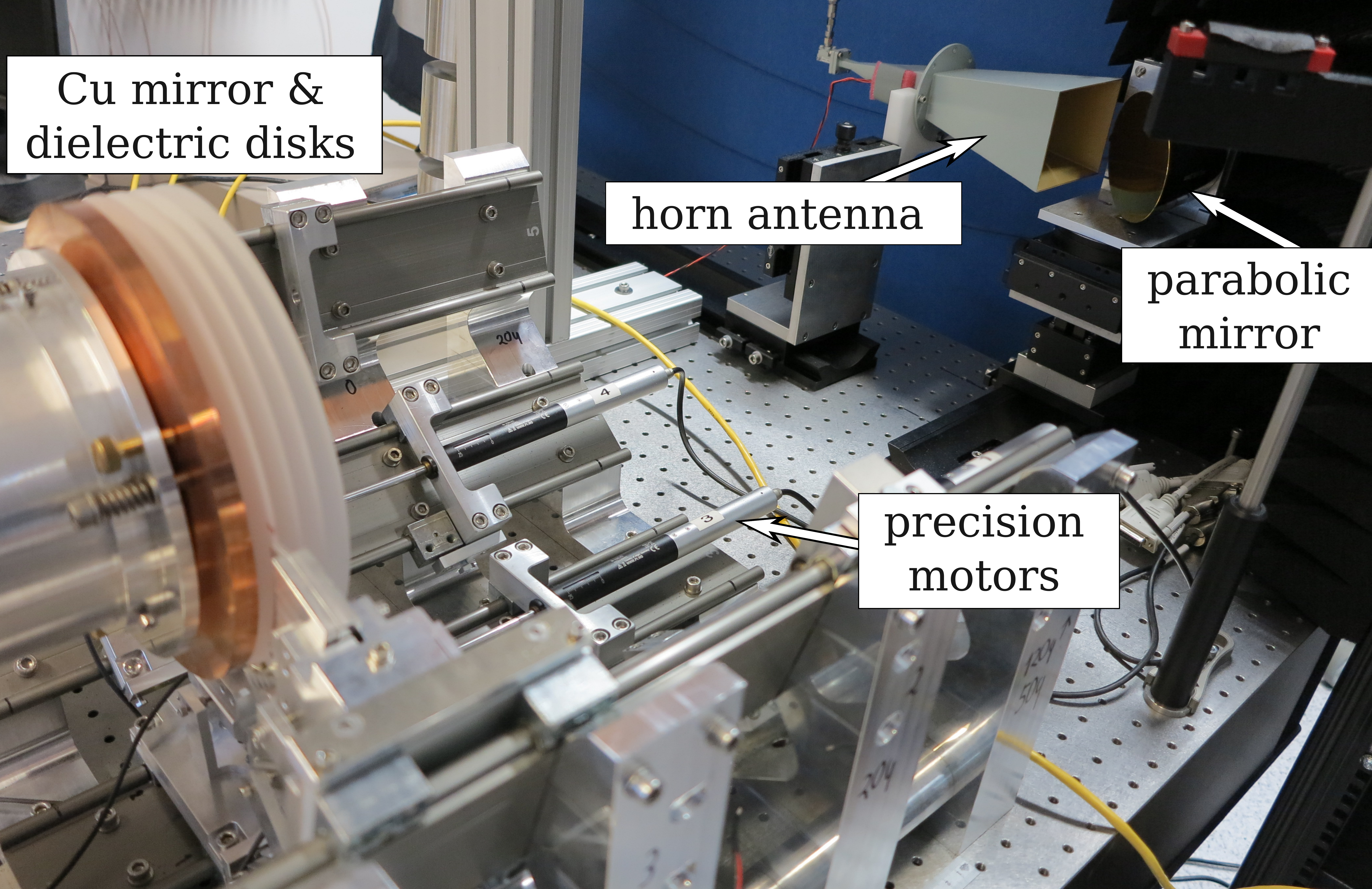}
\includegraphics[width=0.5\textwidth]{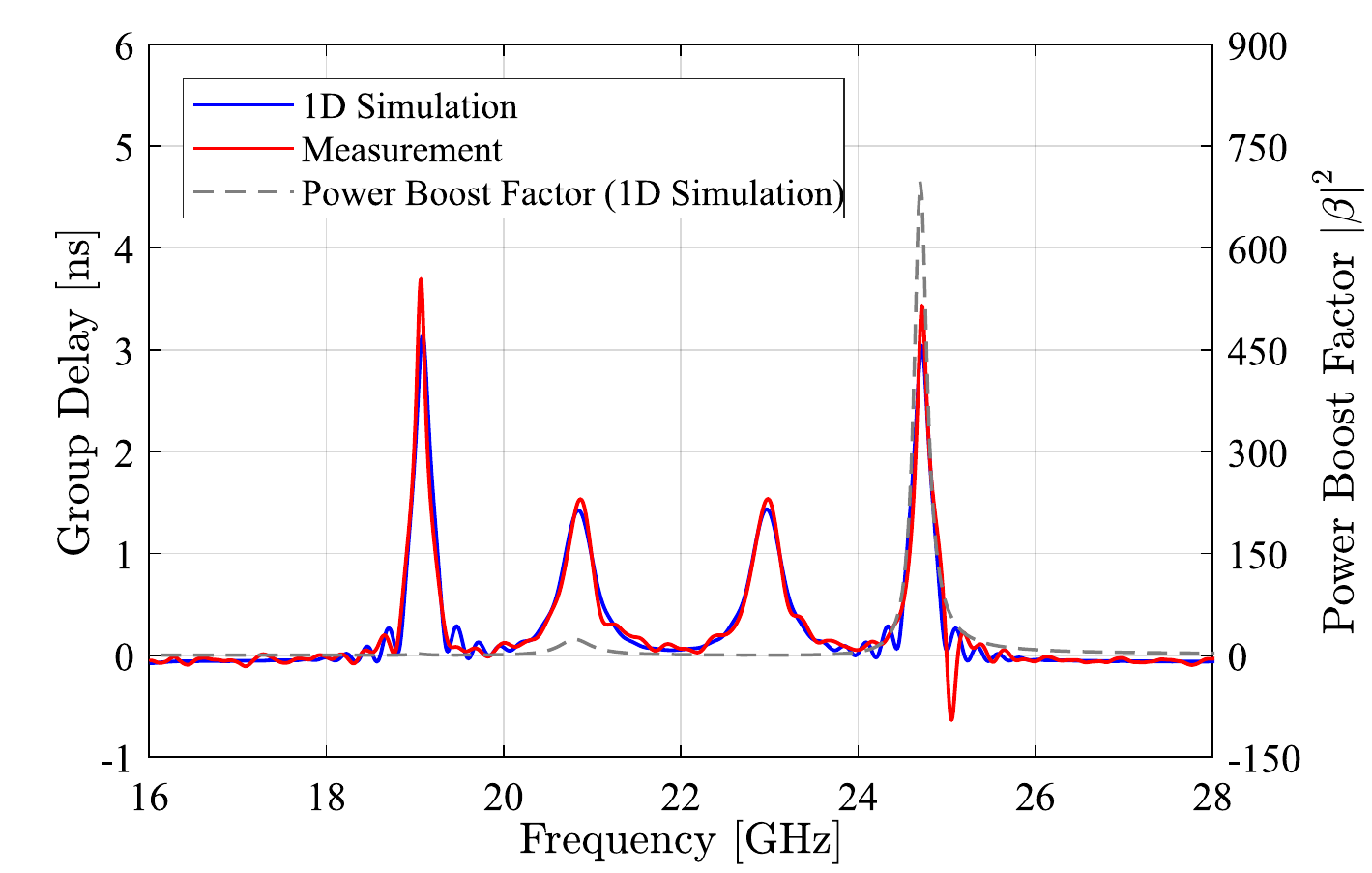}
}
\caption{\textbf{Left:} Prototype booster with 4 disks installed.
\textbf{Right:} Simulated (blue) and measured (red) group delays for 
 a typical 4 disk setup after optimization of disk spacings. The corresponding power boost factor is shown as gray dashed line.
}\label{Fig:4_disk_exemplary}
\end{figure}

As outlined in previous sections, the power boost factor $\beta^2$ can be calculated, considering the axion-induced emitted $E$-field amplitude from each individual disk and the propagation of each of these signals throughout the booster~\cite{foundations}. This implies that transmissivity and reflectivity of the haloscope are correlated to the boost factor curve. As the boost factor cannot be measured directly, the correlation with the group delay, transmissivity and reflectivity is exploited. 
Hence, transmissivity and reflectivity can be used to verify the simulated boost factor behavior  and support the disk placement procedure during the actual operation of the experiment.
 In an ideal loss-less booster the magnitude of the reflectivity will always be unity.
Therefore, it is more feasible to consider its phase or the group delay $\tau_g = - d \Phi / d \omega$, with $\Phi$ the phase of the reflected signal and $\omega$ the angular frequency. The group delay can be qualitatively understood as the mean retention time of reflected photons within the booster, mapping out resonances. The  correlation between boost factor and  group delay is shown in figure~\ref{Fig:4_disk_exemplary}~(right) for a set of four equally spaced disks at \SI{7}{\milli\metre} distances. In such a case it manifests predominantly in the correlation to the group delay peak at the highest frequency.

In order to show that a booster with the predicted electromagnetic properties can actually be built and to study effects that not accounted for in the idealized 1D calculations, the prototype setup shown in figure~\ref{Fig:4_disk_exemplary}~(left) has been developed. It consists of up to five sapphire disks placed in front of a copper mirror. The disks have a dielectric constant of $\epsilon \approx 9.4$ perpendicular to the beam axis, a thickness of \SI{1}{\milli\metre}, and a diameter of \SI{200}{\milli\metre}. Precision motors change their position with an accuracy of a few ${\mu \rm m}$ after accounting for temperature effects and mechanical hysteresis.
The phase and amplitude of a reflected signal at a given frequency can be measured using a Vector Network Analyzer (VNA) connected to the antenna shown in figure~\ref{Fig:4_disk_exemplary}.

The temporal stability was investigated using long term ($\sim 15$ hours) reflectivity measurements with up to 4 installed disks without any disk movement. 
The frequency stability of a single observed group delay peak was on the order of $\pm\,\SI{2}{\mega\hertz}$. This corresponds to an uncertainty in disk position of less than a micrometer. The variations are attributed to mechanical vibration or thermal contraction in the setup. Compared to the envisioned boost factor bandwidth of \SI{50}{\mega\hertz}, this indicates that long-term stability of the system is sufficient even under without active vibrational damping and with temperature fluctuations of the order $\pm \SI{1}{\degreeCelsius}$.

\begin{figure}[tb]
\centering
\includegraphics[width=0.48\linewidth]{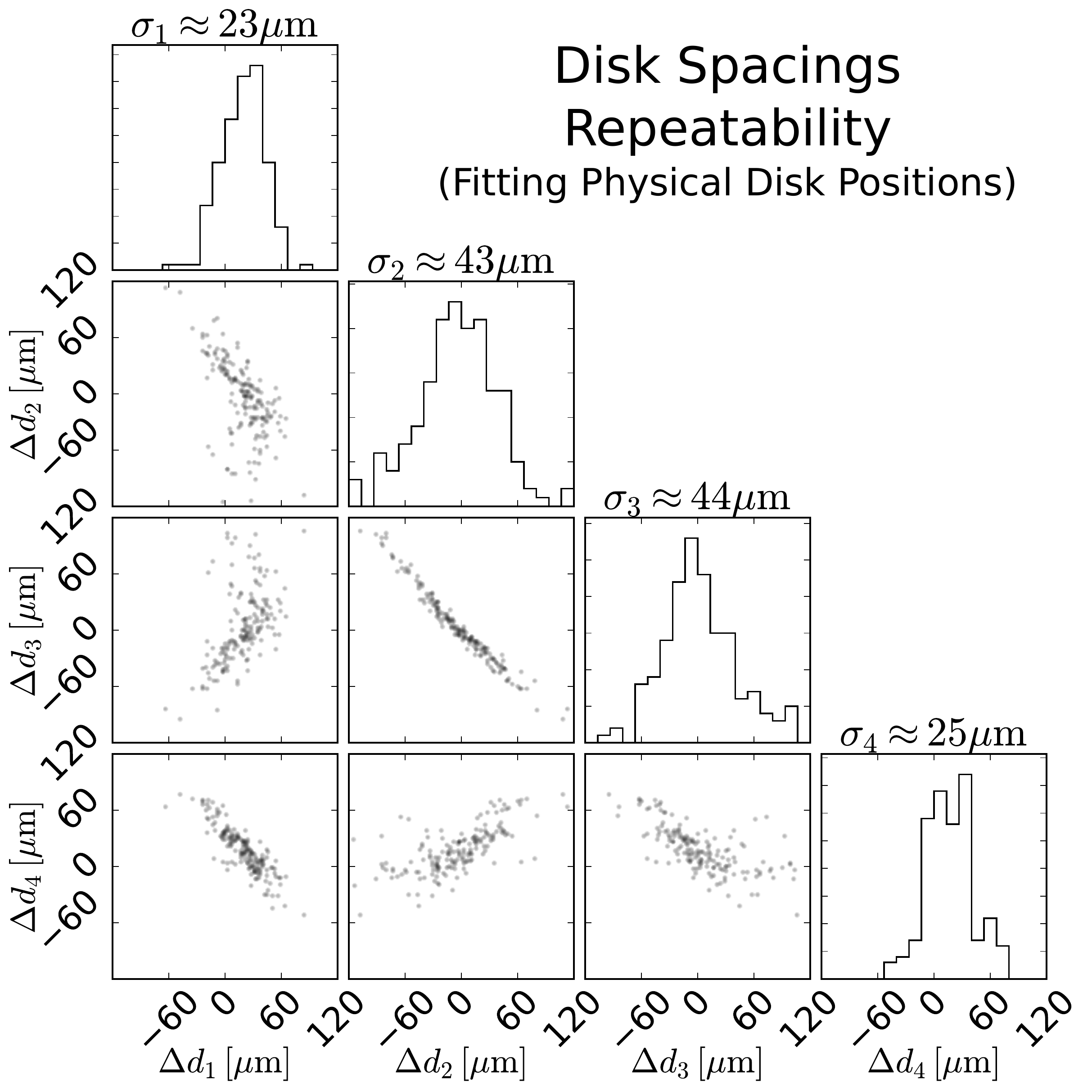}
\includegraphics[width=0.48\linewidth]{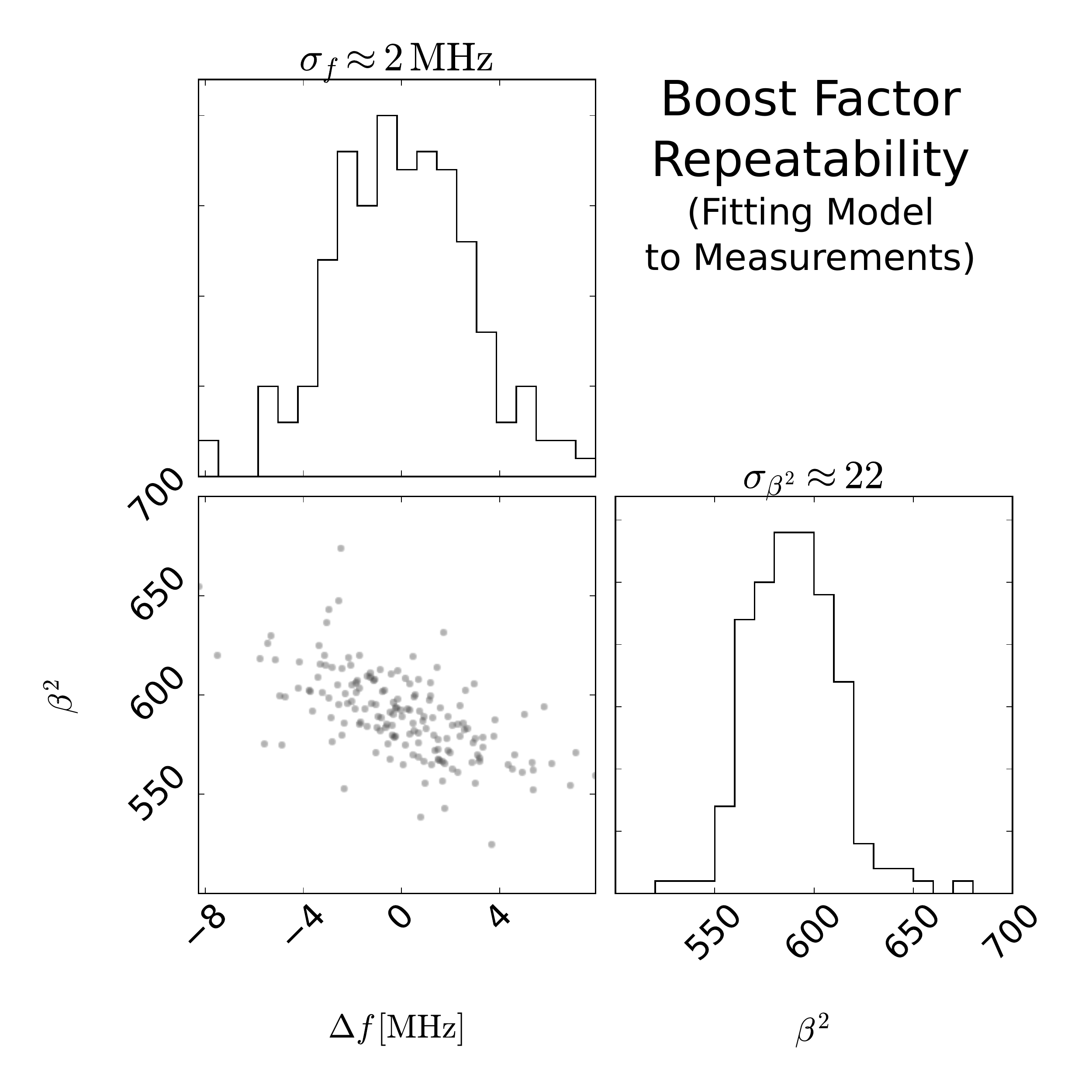}
\caption[Boost Factor Reproducability]{Reproducibility of fit results. \textbf{Left:} Motor position offsets $d_i$ after fitting the physical disk positions such that the measured group delay matches the simulation. \textbf{Right:} Boost factor amplitude and position variation after fitting the model to each of the measured group delays corresponding to the data points in the left plot.}
\label{fig:seed-setup-reproducabilities}
\end{figure}

The configuration of physical disk positions is found by matching the group delay of the system to the prediction of the 1D model.
To this end optimization schemes need to be employed, minimizing the number of function evaluations -- corresponding to a realignment of the setup and measurement of the electromagnetic response -- while avoiding convergence in local minima.
Simple implementations of a genetic \cite{genetic-algorithm} and Nelder-Mead \cite{nelder-mead} algorithm have been tested. The convergence criterion requires changes of the motor positions of less than a $\si{\micro\metre}$.
Figure~\ref{Fig:4_disk_exemplary}~(right) shows a typical group delay fit for a case of 4 equidistant disks at \SI{7}{mm} distance. 
For disk spacings of \SI{8}{\milli\metre}  the situation corresponds to a maximal power boost of $\sim 700$ at a frequency of $\approx \SI{22.4}{\giga\hertz}$ with a FWHM of $\sim \SI{180}{\mega\hertz}$.
While the Nelder-Mead algorithm converges significantly faster, the genetic algorithm seems to be less prone to converge to local minima. 
Figure~\ref{fig:seed-setup-reproducabilities} shows the distribution of motor positions after repeating the optimization multiple times. While the repeatability in single spacings is better than $\sim \SI{40}{\micro\metre}$, the electromagnetic response is degenerate in the disk spacings, such that higher disk separations at the one end can be partially compensated by smaller spacings on the other end of the booster.
To infer the effect on the power boost factor, the disk positions in the model are optimized such that the model predicts the various measured group delays. The change in frequency and power boost are outlined in figure~\ref{fig:seed-setup-reproducabilities}, being $\Delta \beta^2_{\rm max} \approx \SI{10}{\percent}\,\beta^2_{\rm max}$ and $\Delta f \approx \SI{10}{\percent}\,{\rm FWHM}$. We conclude that our system can be aligned to match the predicted response and achieve the desired boost factors within acceptable accuracy for up to 5 disks at equidistant spacings of 6 to 9 mm. A more detailed paper on this setup is in preparation.

\section{Towards realization of MADMAX}
\subsection{Prototype}
The receiver sensitivity has been verified experimentally as discussed in section~\ref{sec:receivertest}, while a basic test of the precision of disk placement has been performed for a setup with four disks of 20\,cm diameter and a mirror as discussed in section~\ref{sec:boostertest}. 
Before a full scale experiment can be built, many components still need to be investigated and developed. The full R\&D programme of the coming years is beyond the scope of this paper. Some of the ongoing investigations include the following:
\begin{itemize}
\item All proposed components have to be tested for their performance in a strong magnetic field of up to \MagneticField and at cryogenic temperatures of \TActualBoosterTemp.
\item  For the final size of the experiment the required accuracy  of relevant parameters (precision of disk positioning, disk thickness, surface roughness, loss mechanisms) need to be tested against available technology.
\item A technology to produce thin dielectric disks with plane surfaces with areas up to \MagnetBore has to be developed and tested, e.g., disk tiling.
\item Procedures to measure the properties of dielectric disks such as permittivity and loss factor need to be adapted. 
\item A  concept to position the disks inside a strong magnetic field in vacuum and at cryogenic temperatures down to \TActualBoosterTemp\ has to be developed.
\item The system noise behavior has to be studied down to temperatures around \TActualBoosterTemp, including the noise behaviour of the booster itself and its surroundings (tube, holding structure). 
\item The simulation of the experiment needs to be tested for realism up to the full scale of the experimental device. The preliminary 1D calculations need to be extended to realistic full 3D simulations.
\item The scaling behavior of the power boost factor $\beta^2$ with the number of disks and their area has to be studied further. The area law (section \ref{sec:foundations}) holds in 1D, while effects such as diffraction, dielectric loss and others may modify the behavior. 
\item The beam shape of the system needs to be studied, and a suitably matched antenna system needs to be devised.
\item The  receiver technology for low frequencies, based on heterodyne mixing needs to be optimized, while new technologies for high frequencies need to be explored.
\item The feasibility of manufacturing a dipole magnet fulfilling the requirements of the experiment needs to be studied in terms of cost and availability of technology.
\end{itemize}

The central component towards accomplishing these goals and to obtain a better understanding of the fundamental concept of the experiment will be a prototype system with smaller dimensions. The prototype setup will have a smaller number of disks ($\approx 20$) each with a smaller diameter ($\approx 30$ cm) than in the full scale experiment. 

This prototype booster will be used to study adjustment of disk positions for a defined frequency and bandwidth. 
The cryogenic performance of the mechanical parts will be studied in a dedicated cryostat that allows the components to be cooled down to \TActualBoosterTemp. It is also foreseen to test the protoype booster inside  magnetic fields of a few T. It is presently being investigated whether a suitable magnet could be available.

The prototype represents a stepping stone towards the development of technologies to be used in the full experiment. Beyond that, the prototype will also provide competitive physics results, covering regions in the axion DM phase space that have not been probed experimentally before, if a suitable magnet can be found.

 As shown in figure \ref{fig:sensitivity}, with a setup consisting of 5 disks with 30\,cm diameter in front of a mirror inside a 3\,T magnetic field, a sensitivity could be achieved that can probe uncovered regions of DM parameter space, and which exceeds the projected sensitivity of the proposed IAXO experiment for solar axions \cite{iaxo}, which does not depend on cosmological assumptions.  
A setup with 20 disks  improves the sensitivity on $g_{a\gamma}$ a factor of two compared to a 5 disk setup. 

 As shown in figure \ref{fig:sensitivity_hp}, even without a magnet, a prototype setup could considerably improve the existing limits in the search for DM hidden photons and probe into unexplored territory in the allowed parameter space. Even a five disk setup would improve the limit obtained in reference \cite{tokyo_limits} for $\sim\,50\,\mu$eV hidden photons  by more than two orders of magnitude.

\subsection{Experimental site}
DESY Hamburg has offered the MADMAX collaboration to host the experiment at
the HERA north hall.  
It has been verified to be a suitable site for the experiment: It fulfills the space and infrastructural requirements and is $\sim$\,20\,m below the surface. It has enough overburden to sufficiently shield 
electromagnetic radiation present at the surface. First background measurements at the site in the frequency range between few\,kHz and $\sim$\,20\,GHz have been performed. Above 1\,GHz the measured background consistently had a level $< -100\,{\rm dBm}$; no signal could be detected. Between 100\,MHz and 1\,GHz some signals from radio stations, telecommunication and emergency communication transmitters are present, but also in this frequency range the general noise was consistently less than $-100$ dBm. No correlations with RF activities of the PETRA and FLASH RF infrastructure---ongoing during the measurements---were found. 
The cryogenic infrastructure necessary for the magnet and booster operation already exists and needs to be adapted. Furthermore, the iron yoke of the H1 detector is still in the hall and could be used as yoke for the MADMAX magnet.

\subsection{Timeline}
As first steps it is planned to answer scientific and strategically important questions that have the potential to influence the design of the experiment. They will be addressed within the next 3--4 years using the  prototype setup described above.
Also a demonstrator magnet is planned that will prove the feasibility of the technology.
Already during this time, physics results will be achieved using the prototype cryogenic booster.

Based on the results from the prototype studies a final design will be devised.
It is envisioned to start building the final experiment once results from the magnet demonstrator setup and MADMAX prototype experiment are available, approximately in the year $\sim$\,2022.
The time scale to start data taking is mainly driven by the production of the magnet. 

The plan outlined in this paper would allow to probe the axion DM mass range between $\sim$\,40 and 120\,$\mu$eV. Until then   new detection techniques with quantum limited detectors would be developed to allow the sensitivity range of the experiment to be extended up to an axion mass of~$\,\sim$\,400\,$\mu$eV.

\section{Conclusions}
Axions are very well motivated particle candidates that can explain both the strong CP problem and the DM problem simultaneously. Their coupling to photons through the Primakoff effect makes them detectable in principle in the laboratory. This has been and is being exploited in experiments relying on the concept of resonant axion to photon conversion in cavities. These experiments are sensitive enough to probe axions as DM candidates in the axion mass range below $\sim$\,40\,$\mu$eV. The axion mass range above 40\,$\mu$eV, which is predicted by theoretical models where the PQ symmetry breaking occurs after inflation, is not yet experimentally explored. 

We propose a new experiment, MADMAX,  a dielectric haloscope.
We have shown that simulations and first experimental tests of the concept are very promising. 
This lead us to the conclusion that this approach is experimentally scalable.
We introduced the baseline design of the MADMAX experiment. This experiment consists of three main components: 
\begin{enumerate}
\item A booster with $\sim$80 disks with area $\sim$\,1\,m$^2$ made from a material with high dielectric constant and low dielectric loss in front of a mirror. The distances between the disks need to be adjustable in a range from  $2-20$ mm with a precision of a few $\mu$m.
\item A magnet with an aperture to host the booster. The magnetic field parallel to the disks along with the size of the disk area needs to achieve a $B^2A$
 value of \MagnetBsquareA.
\item Receivers that can detect microwaves with a power of $\sim$\,10$^{-22}$\,W in a few days of measurement time in the range $10-100\,{\rm GHz}$.
\end{enumerate}
The whole setup needs to be installed in an environment with low electromagnetic noise. Presently the HERA hall north at DESY, Hamburg has been identified as a site.

In the next 2--3 years a  smaller prototype  with disk diameters of $\sim$\,30\,cm will be designed and produced. This will allow  to verify the scalability of the technologies investigated so far. Such a setup would probe uncovered parameter spcace and provide competitive results in the search for hidden photons and also for ALPs if a suitable magnet can be identified.

Once the prototype is commissioned, the construction of the final experiment will be envisaged, taking the experience with the prototype into account. In case of a smooth implementation, first measurements with a sensitivity high enough to probe DM QCD axions could be taken starting from $\sim\,2025$. It would need roughly ten years to probe the whole mass range between 40 and 400 $\mu$eV predicted for axion DM in the case of PQ symmetry breaking after inflation. This proposal presents the so far only known approach to cover this very well motivated mass range for axions and is complementary to other axion DM searches discussed in the literature.

\section{Acknowledgments}
This work is supported by the Deutsche Forschungsgemeinschaft under
Germany's Excellence Strategy - EXC 2121 Quantum Universe -
390833306. Alexander Schmidt is supported by DFG through project number 441532750. Chang Lee is supported by DFG through SFB 1258.


\begin{thebibliography}{99}

\bibitem{Peccei:1977hh}
  R.~D.~Peccei and H.~R.~Quinn,
  %``CP Conservation in the Presence of Instantons,''
  Phys.\ Rev.\ Lett.\  {\bf 38} (1977) 1440.
 % doi:10.1103/PhysRevLett.38.1440
  %%CITATION = doi:10.1103/PhysRevLett.38.1440;%%
  %3805 citations counted in INSPIRE as of 07 Feb 2017
  
\bibitem{Peccei:1977ur} R.~D.~Peccei and H.~R.~Quinn,  Phys.\ Rev.\ D {\bf 16} (1977) 1791.

\bibitem{Weinberg:1977ma}
  S.~Weinberg,
  %``A New Light Boson?,''
  Phys.\ Rev.\ Lett.\  {\bf 40} (1978) 223.
%  doi:10.1103/PhysRevLett.40.223
  %%CITATION = doi:10.1103/PhysRevLett.40.223;%%
  
  \bibitem{Wilczek:1977pj}
  F.~Wilczek,
  %``Problem of Strong  $P$  and  $T$  Invariance in the Presence of Instantons,''
  Phys.\ Rev.\ Lett.\  {\bf 40} (1978) 279.
%  doi:10.1103/PhysRevLett.40.279
  %%CITATION = doi:10.1103/PhysRevLett.40.279;%%

\bibitem{Preskill:1982cy}
  J.~Preskill, M.~B.~Wise and F.~Wilczek,
  %``Cosmology of the Invisible Axion,''
  Phys.\ Lett.\ B {\bf 120} (1983) 127.
%   [Phys.\ Lett.\  {\bf 120B} (1983) 127].
%  doi:10.1016/0370-2693(83)90637-8
  %%CITATION = doi:10.1016/0370-2693(83)90637-8;%%

\bibitem{Abbott:1982af}
  L.~F.~Abbott and P.~Sikivie,
  %``A Cosmological Bound on the Invisible Axion,''
  Phys.\ Lett.\ B {\bf 120} (1983) 133.
%   [Phys.\ Lett.\  {\bf 120B} (1983) 133].
%  doi:10.1016/0370-2693(83)90638-X
  %%CITATION = doi:10.1016/0370-2693(83)90638-X;%%

\bibitem{Dine:1982ah}
  M.~Dine and W.~Fischler,
  %``The Not So Harmless Axion,''
  Phys.\ Lett.\ B {\bf 120} (1983) 137.
%   [Phys.\ Lett.\  {\bf 120B} (1983) 137].
%  doi:10.1016/0370-2693(83)90639-1
  %%CITATION = doi:10.1016/0370-2693(83)90639-1;%%

\bibitem{Sikivie:2006ni}
  P.~Sikivie,
  %``Axion Cosmology,''
  Lect.\ Notes Phys.\  {\bf 741} (2008) 19
%  doi:10.1007/978-3-540-73518-2
  [astro-ph/0610440].
  %%CITATION = doi:10.1007/978-3-540-73518-2_2;%%
  %290 citations counted in INSPIRE as of 14 Sep 2018
  
\bibitem{Marsh:2015xka}
  D.~J.~E.~Marsh,
  %``Axion Cosmology,''
  Phys.\ Rept.\  {\bf 643} (2016) 1
%  doi:10.1016/j.physrep.2016.06.005
  [arXiv:1510.07633].
  %%CITATION = doi:10.1016/j.physrep.2016.06.005;%%
  %294 citations counted in INSPIRE as of 14 Sep 2018

\bibitem{raffelt} G.~Raffelt, Lecture Notes Phys. \textbf{741} (2008) 51

\bibitem{Chang:2018rso}
  J.~H.~Chang, R.~Essig and S.~D.~McDermott, 
  JHEP {\bf 1809} (2018) 051.
%  doi:10.1007/JHEP09(2018)051
  [arXiv:1803.00993].
  
\bibitem{Giannotti:2017hny}
  M.~Giannotti, I.~G.~Irastorza, J.~Redondo, A.~Ringwald and K.~Saikawa,
  %``Stellar Recipes for Axion Hunters,''
  JCAP {\bf 1710} (2017) no.10,  010
%  doi:10.1088/1475-7516/2017/10/010
  [arXiv:1708.02111].
  %%CITATION = doi:10.1088/1475-7516/2017/10/010;%%
  %24 citations counted in INSPIRE as of 08 Nov 2018

\bibitem{RingwaldPDG} A. Ringwald, L. Rosenberg and G. Rybka, M.~Tanabashi {\it et al.} [Particle Data Group],
  %``Review of Particle Physics,''
  Phys.\ Rev.\ D {\bf 98} (2018) no.3,  030001.
%  doi:10.1103/PhysRevD.98.030001
  %%CITATION = doi:10.1103/PhysRevD.98.030001;%%

\bibitem{sikivie} P.~Sikivie, Phys.~Rev.~Lett.~\textbf{51} (1983) 1415; Erratum {\it ibid.} \textbf{51} (1983) 695.
  
\bibitem{admx} S.~Asztalos et al., Phys.~Rev.~D \textbf{69} (2004) 011101.

\bibitem{Du:2018uak} 
  N.~Du {\it et al.} [ADMX Collaboration],
  %``A Search for Invisible Axion Dark Matter with the Axion Dark Matter Experiment,''
  Phys.\ Rev.\ Lett.\  {\bf 120}, no. 15, 151301 (2018)
%  doi:10.1103/PhysRevLett.120.151301
  [arXiv:1804.05750].
  %%CITATION = doi:10.1103/PhysRevLett.120.151301;%%
  %35 citations counted in INSPIRE as of 20 Nov 2018 

\bibitem{McAllister:2017lkb}
  B.~T.~McAllister, G.~Flower, E.~N.~Ivanov, M.~Goryachev, J.~Bourhill and M.~E.~Tobar,
  %``The ORGAN Experiment: An axion haloscope above 15 GHz,''
  Phys.\ Dark Univ.\  {\bf 18} (2017) 67
%  doi:10.1016/j.dark.2017.09.010
  [arXiv:1706.00209].
  %%CITATION = doi:10.1016/j.dark.2017.09.010;%%
  %28 citations counted in INSPIRE as of 20 Nov 2018

\bibitem{admx-hf}
  B.~M.~Brubaker {\it et al.},
  %``First results from a microwave cavity axion search at 24 $\mu$eV,''
  Phys.\ Rev.\ Lett.\  {\bf 118} (2017) no.6,  061302
%  doi:10.1103/PhysRevLett.118.061302
  [arXiv:1610.02580].
  %%CITATION = doi:10.1103/PhysRevLett.118.061302;%%

\bibitem{Kenany:2016tta}
  S.~Al Kenany {\it et al.},
  %``Design and operational experience of a microwave cavity axion detector for the 20-100 $\mu$eV range,''
  Nucl.\ Instrum.\ Meth.\ A {\bf 854} (2017) 11
%  doi:10.1016/j.nima.2017.02.012
  [arXiv:1611.07123].
  %%CITATION = doi:10.1016/j.nima.2017.02.012;%%
  %25 citations counted in INSPIRE as of 14 Sep 2018
  %\cite{Zhong:2018rsr}

\bibitem{Zhong:2018rsr}
  L.~Zhong {\it et al.} [HAYSTAC Collaboration],
  %``Results from phase 1 of the HAYSTAC microwave cavity axion experiment,''
  Phys.\ Rev.\ D {\bf 97} (2018) no.9,  092001
%  doi:10.1103/PhysRevD.97.092001
  [arXiv:1803.03690].
  %%CITATION = doi:10.1103/PhysRevD.97.092001;%%
  %4 citations counted in INSPIRE as of 14 Sep 2018

\bibitem{CULTASK}   W.~Chung, Launching axion experiment at CAPP/IBS in Korea, in {\em Proceedings of the 12th Patras Workshop on Axions, WIMPs and WISPs, Jeju, Korea } (2016).

\bibitem{1}
  M.~Kawasaki, K.~Saikawa and T.~Sekiguchi,
  %``Axion dark matter from topological defects,''
  Phys.\ Rev.\ D {\bf 91} (2015) no.6,  065014
%  doi:10.1103/PhysRevD.91.065014
  [arXiv:1412.0789].
  %%CITATION = doi:10.1103/PhysRevD.91.065014;%%

\bibitem{2} T. Hiramatsu, M. Kawasaki, K. Saikawa and T. Sekiguchi, Phys. Rev. D \textbf{85} (2012) 105020, Erratum ibid. \textbf{86} (2012) 089902.

\bibitem{3} E. W. Kolb and I. I. Tkachev, Phys. Rev. D \textbf{49} (1994) 5040.

\bibitem{4} K. M. Zurek, C. J. Hogan and T. R. Quinn, Phys. Rev. D \textbf{75} (2007) 043511.

\bibitem{nature} S. Borsanyi et al., Nature \textbf{539} (2016) 69.

\bibitem{smash}
  G.~Ballesteros, J.~Redondo, A.~Ringwald and C.~Tamarit,
  %``Standard Model-axion-seesaw-Higgs portal inflation. Five problems of particle physics and cosmology solved in one stroke,''
  JCAP {\bf 1708} (2017) no.08,  001
%  doi:10.1088/1475-7516/2017/08/001
  [arXiv:1610.01639].
  %%CITATION = doi:10.1088/1475-7516/2017/08/001;%%

\bibitem{Klaer:2017ond}
  V.~B.~Klaer and G.~D.~Moore,
  %``The dark-matter axion mass,''
  JCAP {\bf 1711} (2017) no.11,  049
%  doi:10.1088/1475-7516/2017/11/049
  [arXiv:1708.07521].
  %%CITATION = doi:10.1088/1475-7516/2017/11/049;%%
  %32 citations counted in INSPIRE as of 14 Sep 2018

\bibitem{Gorghetto:2018myk}
  M.~Gorghetto, E.~Hardy and G.~Villadoro,
  %``Axions from Strings: the Attractive Solution,''
  JHEP {\bf 1807} (2018) 151
%  doi:10.1007/JHEP07(2018)151
  [arXiv:1806.04677].
  %%CITATION = doi:10.1007/JHEP07(2018)151;%%
  %11 citations counted in INSPIRE as of 14 Sep 2018

\bibitem{prl}
  A.~Caldwell {\it et al.} [MADMAX Working Group],
  %``Dielectric Haloscopes: A New Way to Detect Axion Dark Matter,''
  Phys.\ Rev.\ Lett.\  {\bf 118} (2017) no.9,  091801
%  doi:10.1103/PhysRevLett.118.091801
  [arXiv:1611.05865].
  %%CITATION = doi:10.1103/PhysRevLett.118.091801;%%

\bibitem{dish} D. Horns, J. Jaeckel, A. Lindner, A. Lobanov, J. Redondo and A. Ringwald, 
%Searching for WISPy Cold Dark Matter with a Dish Antenna,
 JCAP \textbf{1304}(2013)016 [arXiv:1212.2970].

\bibitem{dielectric} J.~Jaeckel and J.~Redondo, Phys.\ Rev.\ D \textbf{88} (2013) 115002 [arXiv:1308.1103].

\bibitem{Pospelov:1999mv}
  M.~Pospelov and A.~Ritz,
  %``Theta vacua, QCD sum rules, and the neutron electric dipole moment,''
  Nucl.\ Phys.\ B {\bf 573} (2000) 177
%  doi:10.1016/S0550-3213(99)00817-2
  [hep-ph/9908508].
  %%CITATION = doi:10.1016/S0550-3213(99)00817-2;%%
  %75 citations counted in INSPIRE as of 14 Sep 2018

\bibitem{Baker:2006ts}
  C.~A.~Baker {\it et al.},
  %``An Improved experimental limit on the electric dipole moment of the neutron,''
  Phys.\ Rev.\ Lett.\  {\bf 97} (2006) 131801
 % doi:10.1103/PhysRevLett.97.131801
  [hep-ex/0602020].
  %%CITATION = doi:10.1103/PhysRevLett.97.131801;%%
  %889 citations counted in INSPIRE as of 07 Feb 2017

\bibitem{Pospelov:2005pr} M.~Pospelov and A.~Ritz,  Annals Phys.\  {\bf 318} (2005) 119 [hep-ph/0504231].

\bibitem{Vafa:1984xg}  C.~Vafa and E.~Witten,
  Phys.\ Rev.\ Lett.\  {\bf 53} (1984) 535.
    
\bibitem{diCortona:2015ldu}
  G.~Grilli di Cortona, E.~Hardy, J.~Pardo Vega and G.~Villadoro,
  %``The QCD axion, precisely,''
  JHEP {\bf 1601} (2016) 034
  %doi:10.1007/JHEP01(2016)034
  [arXiv:1511.02867].
  %\cite{Giannotti:2015kwo}
  
  
 \bibitem{DiLuzio:2016sbl}
 L.~Di Luzio, F.~Mescia and E.~Nardi,
 %``Redefining the Axion Window,''
 Phys.\ Rev.\ Lett.\  {\bf 118} (2017) no.3,  031801
 doi:10.1103/PhysRevLett.118.031801
 [arXiv:1610.07593 [hep-ph]].
 %%CITATION = doi:10.1103/PhysRevLett.118.031801;%%
 %49 citations counted in INSPIRE as of 16 Jan 2019
%\cite{Agrawal:2017cmd}
\bibitem{Agrawal:2017cmd}
 P.~Agrawal, J.~Fan, M.~Reece and L.~T.~Wang,
 %``Experimental Targets for Photon Couplings of the QCD Axion,''
 JHEP {\bf 1802} (2018) 006
 doi:10.1007/JHEP02(2018)006
 [arXiv:1709.06085 [hep-ph]].
 %%CITATION = doi:10.1007/JHEP02(2018)006;%%
 %20 citations counted in INSPIRE as of 16 Jan 2019

  

\bibitem{Giannotti:2015kwo}
  M.~Giannotti, I.~Irastorza, J.~Redondo and A.~Ringwald,
  %``Cool WISPs for stellar cooling excesses,''
  JCAP {\bf 1605} (2016) no.05,  057
  %doi:10.1088/1475-7516/2016/05/057
  [arXiv:1512.08108].
  %%CITATION = doi:10.1088/1475-7516/2016/05/057;%%
  %11 citations counted in INSPIRE as of 22 Feb 2017

\bibitem{Irastorza:2018dyq}
  I.~G.~Irastorza and J.~Redondo,
  %``New experimental approaches in the search for axion-like particles,''
  Prog.\ Part.\ Nucl.\ Phys.\  {\bf 102} (2018) 89
%  doi:10.1016/j.ppnp.2018.05.003
  [arXiv:1801.08127].
  %\cite{Sikivie:2006ni}

\bibitem{Sikivie:1982qv}
  P.~Sikivie,
  %``Of Axions, Domain Walls and the Early Universe,''
  Phys.\ Rev.\ Lett.\  {\bf 48} (1982) 1156.
%  doi:10.1103/PhysRevLett.48.1156
  %%CITATION = doi:10.1103/PhysRevLett.48.1156;%%
  %544 citations counted in INSPIRE as of 14 Sep 2018
 
\bibitem{Daido:2017wwb} 
  R.~Daido, F.~Takahashi and W.~Yin,
  %``The ALP miracle: unified inflaton and dark matter,''
  arXiv:1702.03284.
  %%CITATION = ARXIV:1702.03284;%%
  
  \bibitem{Ringwald:2015dsf} 
  A.~Ringwald and K.~Saikawa,
% Axion dark matter in the post-inflationary Peccei-Quinn symmetry breaking scenario,
  Phys.\ Rev.\ D {\bf 93}, 085031 (2016);
  Addendum {\em ibid.} {\bf 94}, 049908 (2016)
%  %doi:10.1103/PhysRevD.93.085031, 10.1103/PhysRevD.94.049908
  [arXiv:1512.06436].

\bibitem{Fleury:2015aca}
  L.~Fleury and G.~D.~Moore,
  %``Axion dark matter: strings and their cores,''
  JCAP {\bf 1601} (2016) 004
 % doi:10.1088/1475-7516/2016/01/004
  [arXiv:1509.00026].
  %%CITATION = doi:10.1088/1475-7516/2016/01/004;%%
  %9 citations counted in INSPIRE as of 21 Feb 2017

\bibitem{Fleury:2016xrz}
  L.~M.~Fleury and G.~D.~Moore,
  %``Axion String Dynamics I: 2+1D,''
  JCAP {\bf 1605} (2016) 005
 % doi:10.1088/1475-7516/2016/05/005
  [arXiv:1602.04818].
  %%CITATION = doi:10.1088/1475-7516/2016/05/005;%%
  %6 citations counted in INSPIRE as of 21 Feb 2017
  %\cite{Moore:2016itg}

\bibitem{Moore:2016itg}
  G.~D.~Moore,  arXiv:1604.02356.

\bibitem{Budker:2013hfa}   D.~Budker, P.~W.~Graham, M.~Ledbetter, S.~Rajendran and A.~Sushkov,  Phys.\ Rev.\ X {\bf 4}, (2014) 021030 [arXiv:1306.6089].

\bibitem{Sikivie:2013laa} 
  P.~Sikivie, N.~Sullivan and D.~B.~Tanner,
 % Proposal for axion dark matter detection using an LC Circuit,
  Phys.\ Rev.\ Lett.\  {\bf 112}, (2014) 131301
  %doi:10.1103/PhysRevLett.112.131301
  [arXiv:1310.8545].
  %%CITATION = doi:10.1103/PhysRevLett.112.131301;%%
  %37 citations counted in INSPIRE as of 29 Sep 2016

\bibitem{Kahn:2016aff}
  Y.~Kahn, B.~R.~Safdi and J.~Thaler,
 % Broadband and resonant approaches to axion dark matter detection,
  Phys.\ Rev.\ Lett.\  {\bf 117}, (2016) 141801
  [arXiv:1602.01086].
  %%CITATION = doi:10.1103/PhysRevLett.117.141801;%%
  %6 citations counted in INSPIRE as of 18 Nov 2016

\bibitem{Ouellet:2018beu}
  J.~L.~Ouellet {\it et al.},
  %``First Results from ABRACADABRA-10 cm: A Search for Sub-$\mu$eV Axion Dark Matter,''
  arXiv:1810.1225.
  %%CITATION = ARXIV:1810.12257;%%
  %7 citations counted in INSPIRE as of 09 Jan 2019

\bibitem{Rybka:2014cya}
  G.~Rybka, A.~Wagner, A.~Brill, K.~Ramos, R.~Percival and K.~Patel,
  %``Search for dark matter axions with the Orpheus experiment,''
  Phys.\ Rev.\ D {\bf 91} (2015) no.1,  011701
%  doi:10.1103/PhysRevD.91.011701
  [arXiv:1403.3121].
  %%CITATION = doi:10.1103/PhysRevD.91.011701;%%
  %63 citations counted in INSPIRE as of 11 Jan 2019

\bibitem{Melcon:2018dba}
  A.~A.~Melcon {\it et al.},
  %``Axion Searches with Microwave Filters: the RADES project,''
  JCAP {\bf 1805} (2018) no.05,  040
%  doi:10.1088/1475-7516/2018/05/040
  [arXiv:1803.01243].
  %%CITATION = doi:10.1088/1475-7516/2018/05/040;%%
  %8 citations counted in INSPIRE as of 11 Jan 2019

\bibitem{Silva-Feaver:2016qhh}
  M.~Silva-Feaver {\it et al.},
  %``Design Overview of DM Radio Pathfinder Experiment,''
  IEEE Trans.\ Appl.\ Supercond.\  {\bf 27} (2017) no.4,  1400204
%  doi:10.1109/TASC.2016.2631425
  [arXiv:1610.09344].
  %%CITATION = doi:10.1109/TASC.2016.2631425;%%
  %20 citations counted in INSPIRE as of 09 Jan 2019

\bibitem{Arvanitaki:2014dfa} 
  A.~Arvanitaki and A.~A.~Geraci,
 % Resonantly detecting axion-mediated forces with nuclear magnetic resonance,
  Phys.\ Rev.\ Lett.\  {\bf 113}, (2014) 161801
 % doi:10.1103/PhysRevLett.113.161801
  [arXiv:1403.1290].
  %%CITATION = doi:10.1103/PhysRevLett.113.161801;%%
  %29 citations counted in INSPIRE as of 28 Sep 2016

\bibitem{foundations} 
%\bibitem{Millar:2016cjp}
  A.~J.~Millar, G.~G.~Raffelt, J.~Redondo and F.~D.~Steffen,
  %``Dielectric Haloscopes to Search for Axion Dark Matter: Theoretical Foundations,''
  JCAP {\bf 1701} (2017) no.01,  061
  %doi:10.1088/1475-7516/2017/01/061
  [arXiv:1612.07057].
  %%CITATION = doi:10.1088/1475-7516/2017/01/061;%%
  %24 citations counted in INSPIRE as of 17 Jan 2019

\bibitem{Ioannisian:2017srr}
  A.~N.~Ioannisian, N.~Kazarian, A.~J.~Millar and G.~G.~Raffelt,
  %``Axion-photon conversion caused by dielectric interfaces: quantum field calculation,''
  JCAP {\bf 1709} (2017) no.09,  005
  %doi:10.1088/1475-7516/2017/09/005
  [arXiv:1707.00701].
  %%CITATION = doi:10.1088/1475-7516/2017/09/005;%%
  %10 citations counted in INSPIRE as of 09 Jan 2019

\bibitem{lalo3_a}  J.~Krupka  et al.,  IEEE Transactions on Microwave Theory and Techniques, \textbf{42} (10) 1994 .

\bibitem{lalo3_b} T.~Shimada et al., IEEE Transactions on Ultrasonics, Ferroelectrics, and Frequency Control, \textbf{57} (10) 2010.

\bibitem{heterodyne} T. Oxley, IEEE Trans. Microwave Theo. and Techn., \textbf{50} (3) (2002) 867.

\bibitem{hemt} T. Mimura, IEEE Trans. Microwave theory and techn. \textbf{50} (3) (2002) 780.

\bibitem{dicke} T. Hunter and R. Kimberk, [arXiv:1507.04280].

\bibitem{dicke2} J. Kraus, "Antennas and wave propagation", Mc Graw Hill, (2010), ISBN0070671559.
  
\bibitem{bibinnovationpartner}https://ec.europa.eu/growth/content/8699-innovation-partnerships-keep-public-services-date\_en

\bibitem{cosine}
L. Rossi, %``Superconducting magnets for the LHC main lattice''
IEEE Trans.\ Appl.\ Supercond., vol.~14, no.~2, pp.~153, Jun~2004.

\bibitem{canted} S.~Caspi et al., IEEE Transactions on Applied Superconductivity \textbf{24} (3) (2014) 4001804.

\bibitem{racetrack} E. Rochepault, P. Vedrine and F. Bouillault. IEEE Transactions on Applied Superconductivity \textbf{22} (3) (2012) 4900804.

\bibitem{block} E. Rochepault, P. Vedrine, and F. Bouillault,
IEEE TRANSACTIONS ON APPLIED SUPERCONDUCTIVITY, VOL. 22, NO. 3, JUNE 2012.
  
\bibitem{Millar:2017eoc}
  A.~J.~Millar, J.~Redondo and F.~D.~Steffen,
  %``Dielectric haloscopes: sensitivity to the axion dark matter velocity,''
  JCAP {\bf 1710} (2017) no.10,  006
   Erratum: [JCAP {\bf 1805} (2018) no.05,  E02]
%  doi:10.1088/1475-7516/2017/10/006, 10.1088/1475-7516/2018/05/E02
  [arXiv:1707.04266].
  %%CITATION = doi:10.1088/1475-7516/2017/10/006, 10.1088/1475-7516/2018/05/E02;%%
  %10 citations counted in INSPIRE as of 17 Jan 2019
  
\bibitem{al2o3_a} J.~Krupka et al 1999 Meas. Sci. Technol. \textbf{10} 387.

\bibitem{al2o3_b} X.~Aupi et al., J. Appl. Phys \textbf{95} (5) (2004) 2639.

\bibitem{iaxo} E.~Armengaud et al., JINST \textbf{9} (2014) T05002.

\bibitem{hidden_photons} P.~Arias et al, JCAP \textbf{06} (2012) 013.
  
\bibitem{tokyo_limits} J.~Suzuki, Y.~Inoue, T.~Horie, M. Minowa, Proceedings contribution to the 11th Patras Workshop on Axions, WIMPs and WISPs, Zaragoza, June 22 to 26, 2015, [arXiv:1509.00785].

\bibitem{Jaeckel:2013ija}
  J.~Jaeckel,
  %``A force beyond the Standard Model - Status of the quest for hidden photons,''
  Frascati Phys.\ Ser.\  {\bf 56} (2012) 172
  [arXiv:1303.1821].
  %%CITATION = ARXIV:1303.1821;%%
  %69 citations counted in INSPIRE as of 11 Jan 2019

\bibitem{receiver_statistics_paper}
  F.~Beaujean, A.~Caldwell and O.~Reimann,
  %``Is the bump significant? An axion-search example,''
  Eur.\ Phys.\ J.\ C {\bf 78} (2018) no.9,  793
%  doi:10.1140/epjc/s10052-018-6217-y
  [arXiv:1710.06642].
  %%CITATION = doi:10.1140/epjc/s10052-018-6217-y;%%
  %1 citations counted in INSPIRE as of 05 Nov 2018

\bibitem{genetic-algorithm}
  D.~Whitley, 
  Stat Comput (1994) 4: 65.
%  doi:10.1007/BF00175354
  
\bibitem{nelder-mead}
  J.~C.~Lagarias, J.~A.~Reeds, M.~H.~Wright, and P.~E.~Wright,
  %Convergence Properties of the Nelder--Mead Simplex Method in Low Dimensions,
  SIAM J. Optim., 9(1), 112-147.
%  doi:10.1137/S1052623496303470
  
 
\end{thebibliography}
\end{document}